\documentclass[iop]{emulateapj}
\usepackage{ulem}
\usepackage{color}
\usepackage{ascmac}

\begin{document}

\shorttitle{TERRESTRIAL PLANET FORMATION UNDER THE INFLUENCE OF A HOT JUPITER}
\shortauthors{Ogihara et al.}

\title{\textit{N}--BODY SIMULATIONS OF TERRESTRIAL PLANET FORMATION UNDER THE INFLUENCE OF A HOT JUPITER}

\author{Masahiro Ogihara\altaffilmark{1,}\altaffilmark{2}, Hiroshi Kobayashi\altaffilmark{2} and Shu-ichiro inutsuka\altaffilmark{2}}

\altaffiltext{1}{Observatoire de la C\^ote d'Azur,
Boulevard de l'Observatoire, 06304 Nice Cedex 4, France}
\email{omasahiro@oca.eu}

\altaffiltext{2}{Nagoya University,
Furo-cho, Chikusa-ku, Nagoya, Aichi 464-8602, Japan}

\begin{abstract}
We investigate the formation of multiple--planet systems in the presence of a hot Jupiter
using extended \textit{N}--body simulations that are performed simultaneously with
semi--analytic calculations. Our primary aims are to describe the planet formation process
starting from planetesimals using high--resolution simulations, and to examine the dependences
of the architecture of planetary systems on input parameters (e.g., disk mass, 
disk viscosity). We observe that protoplanets that arise from oligarchic 
growth and undergo type I migration stop migrating when they join a chain of resonant
planets outside the orbit of a hot Jupiter. The formation of a resonant chain is 
almost independent of our model parameters, and is thus a robust process. At the end of 
our simulations, several terrestrial planets remain at around 0.1 AU. 
The formed planets are not equal--mass; the largest planet constitutes 
more than 50 percent of the total mass in the close--in region, which is also less dependent
on parameters. 
In the previous work of this paper \citep{ogihara_etal13}, we have found a new physical mechanism
of induced migration of the hot Jupiter, which is called a crowding--out.
If the hot Jupiter opens up
a wide gap in the disk (e.g., owing to low disk viscosity), crowding--out becomes less efficient
and the hot Jupiter remains.
We also discuss angular momentum transfer between the planets and 
disk.
\end{abstract}

\section{INTRODUCTION}
\label{sec:intro}

Over 900 extrasolar planets have been discovered so far; a large fraction of them are
close--in giant planets (``hot Jupiters'' or HJs) which account for more than 20 percent of all
exoplanets. Because HJs are considered to be gaseous planets, they are formed 
in protoplanetary disks during their formation era, which may
affect the subsequent formation of terrestrial planets. 
For the origin of HJs, there are two commonly--invoked models that include type II migration
(e.g., \citealt{lin_etal96}) and tidal circularization of high--eccentricity planets
(e.g., \citealt{nagasawa_etal08}) based on the standard scenario of planet formation. 
In addition, we introduce a hybrid scenario of planet formation \citep{inutsuka09},
in which  giant planets that formed through gravitational
instability can survive until the accretion phase of terrestrial planets.

Recent resistive magnetohydrodynamic simulations of the formation of protostars and protoplanetary disks show
the formation of multiple planetary--mass objects in the massive circumstellar disks in their formation 
stages (\citealt{inutsuka_etal10}; \citealt{machida_etal10,machida_etal11a,machida_etal11b,machida_etal14}). 
Those objects tend to migrate 
inward rapidly in the early evolutionary phase of disks 
(\citealt{machida_etal11b}; \citealt{baruteau_etal11}). To 
determine the fates of the objects, realistic numerical simulations of long--term evolution of those 
systems are required, but still remains computationally infeasible 
(see, however, \citealt{vorobyov_basu10} for their efforts on 2D simulations without magnetic field). 
On the other hand, recent observations of protoplanetary disks (e.g., \citealt{andrews_etal11}) and 
theoretical work on the disk accretion (e.g., \citealt{suzuki_inutsuka09}; \citealt{suzuki_etal10};
\citealt{fromang_etal13}; \citealt{bai_stone13}) indicate that an inner cavity tends to be 
created in a 
relatively early phase of disk accretion stage, which eventually stops the planetary migration in the 
inner region of the disk. Therefore we can envision that some of gaseous planetary--mass objects 
formed through the gravitational fragmentation of massive disks undergo halfway migration to the inner 
regions and remain as HJs.
This model provides a possible origin of the HJ, in addition to the commonly--invoked models.

Giant planets, such as HJs, gravitationally influence the formation of terrestrial planets
in several ways. To date, the formation of terrestrial planets in the presence of
giant planets has been studied from 
several perspectives (e.g., \citealt{kortenkamp_etal01}; \citealt{levison_agnor03};
\citealt{fogg_nelson07, fogg_nelson09}). For example, \citet{raymond_etal06} performed
\textit{N}--body simulations to investigate the formation of habitable planets during and
after giant planet migration and found that water--rich planets can survive outside the
orbit of giant planets.

Giant planets can open up a ring--like gap in a protoplanetary disk 
(e.g., \citealt{crida_etal06}; \citealt{tanigawa_ikoma07}), outside of which a radial pressure
maximum is created. \citet{ayliffe_etal12} demonstrated with SPH simulations 
that the inward migration of meter--sized solid bodies is efficiently halted at the pressure
maximum, which may trigger gravitational collapse (see also \citealt{lyra_etal09}). \citet{kobayashi_etal12}  
conducted numerical simulations that include collisional fragmentation of solid bodies 
developed by \citet{kobayashi_etal10, kobayashi_etal11} and found that
fragments produced from planetesimals are accumulated at the edge of a Jovian--opened
density gap, leading to the rapid formation of Saturn's core.

Several \textit{N}--body investigations have also been carried out.
\citet{thommes05} considered the case in which a giant planet is located at about 5 AU and 
several planetary cores placed outside its orbit 
undergo type I inward migration. They observed that such planetary 
cores cease their migration by being captured into mean motion resonances (MMRs) with the 
giant planet. Several bodies are in 3:2 or 2:1 MMRs at the end of 
simulation, thus many bodies are in 1:1 commensurabilities with each other.
Planetary cores are not lost via collision with the central star, which may act to enhance 
the growth of planets outside the orbit of the giant planet.

Type I migration can also be halted if the planet is in a region of positive surface density
gradient where the coorbital corotation torque acts as a planet trap \citep{masset_etal06},
which is neglected by \citet{thommes05}. \citet{morbidelli_etal08} calculated the orbital
evolution of several solid planets that undergo type I migration toward the outer edge of
the density gap and confirmed that the planets can survive at the planet trap.
\citet{jakubik_etal12} performed \textit{N}--body investigations of protoplanets ($N$=10--30)
in the presence of Jupiter and Saturn to examine the possibility of the accretion of
Uranus and Neptune outside the gap opened by the giant planets. They found that
more than two planets form at the planet trap, where the most massive planets are
much larger than the second most massive cores.

We perform \textit{N}--body simulations of the accretion of close--in terrestrial planets in the presence
of an HJ. In this study, the growth of protoplanets is calculated using high--resolution
simulations, and the long term evolution for about $10^9$ orbits is examined.
In addition, unlike most previous \textit{N}--body simulations where all bodies that
are handled in the calculation are placed in the initial setup, thus rather limiting the calculation region, 
we use a new, more realistic code in which the \textit{N}--body simulation is combined
with a semianalytical calculation of planet formation in order to consider the migration of
protoplanets from distant regions.
Furthermore, we use several parameters that indicate uncertainties in the planet formation
model (e.g., disk profile, type I migration rate) and vary them over wide ranges to discuss the 
dependences of the results on the parameters.

We especially focus on the close--in region, and thus our results are suitable for comparison
with observational data of exoplanets. Recent observations have discovered multiple 
planetary systems in such regions, and several basic properties have been revealed.
For example, there is a lack of companion planets near the orbit of HJs
(e.g., \citealt{steffen_etal12}). \citet{ogihara_etal13} (hereafter OIK13) investigated
the formation of terrestrial planets outside the orbit of HJs assuming a relatively 
high--viscosity disk and found that the orbit of the HJ moves inward by being pushed by
terrestrial planets that are captured in a 2:1 MMR with the HJ, which is called 
``crowding--out.'' Through this mechanism, we proposed a possible origin for the lack of 
additional planets in HJ systems. In this paper, we also discuss the dependence of the results on disk viscosity.

In the previous letter (OIK13), we proposed a new physical mechanism of 
crowding--out, and this paper extends OIK13 mainly in terms of the following points of view.
(1) By performing high--resolution \textit{N}--body simulations of planetary accretion
from planetesimals, we investigate planet formation along with the growth of protoplanets.
(2) We adopt several model parameters and vary them over wide ranges in order to examine
the dependences of the results on the parameters.
(3) We also carry out in--depth discussions, for example, on angular momentum transfer between
planets.

The structure of this paper is as follows. In Section~\ref{sec:model}, we describe the
numerical methods; in Section~\ref{sec:results1}, we present the results of high--resolution
\textit{N}--body simulations; and in Section~\ref{sec:results2}, we show the results of
\textit{N}--body simulations using model parameters varied over wide ranges.
In Section~\ref{sec:torques}, we analyze the results and examine angular momentum
transfer. In Section~\ref{sec:discussion}, we give a discussion of the parameter dependence of the
results, and in Section~\ref{sec:observation}, we compare our results with
observational properties.
In Section~\ref{sec:conclusions}, we offer our conclusion. 

\section{MODEL DESCRIPTION}
\label{sec:model}
We use the same model as that used in OIK13; a more detailed
description of the model is presented below. 

\subsection{Disk Model}
We investigate a disk model with a power--law radial surface density distribution and
density decay within timescale $t_{\rm dep}$ of
\begin{equation}
\label{eq:surface_gas}
\Sigma_{\rm g} = 2400 f_{\rm g} \left(\frac{r}{1~{\rm AU}}\right)^{-3/2}
\exp\left(\frac{-t}{t_{\rm dep}}\right) {\rm ~g~cm^{-2}},
\end{equation}
where $f_{\rm g}$ and $r$ are a scaling factor and the radial distance from the 
central star, respectively. When $f_{\rm g}=1$ is assumed, 
$\Sigma_{\rm g}$ is 1.4 times that of the minimum-mass solar nebula (MMSN). 
In a series of our simulations, $f_{\rm g}$ is varied between 0.01 and 10 to
explore the result in a starved/massive disk.
The dissipation of a gaseous disk is modeled as an exponential decay 
with the depletion timescale $t_{\rm dep}$; $t_{\rm dep} = 10^6 {\rm yr} = 3 \times 10^7 T_{\rm K}$
is usually used, where $T_{\rm K}$ is the orbital period at 0.1 AU.

The sound speed,  $c_{\rm s} \equiv \sqrt{kT/\mu}$, is 
\begin{equation}
\label{eq:sound_speed}
c_{\rm s} = 1 \times 10^5 \left(\frac{r}{1~{\rm AU}}\right)^{-1/4}
\left(\frac{L_*}{L_\odot}\right)^{1/8} {\rm ~cm~s^{-1}},
\end{equation}
where $k$ is the Boltzmann constant, $\mu$ is the mean molecular weight, and $L_*$
and $L_\odot$ are the luminosities of the host star and the Sun, respectively.
The temperature distribution of an optically thin disk \citep{hayashi81} is:
\begin{equation}
\label{eq:temp}
T = 280 \left(\frac{r}{1~{\rm AU}}\right)^{-1/2} 
\left(\frac{L_*}{L_\odot}\right)^{-1/4} {\rm ~K}.
\end{equation}
Then the disk scale height, $h = \sqrt{2}c_{\rm s}/\Omega_{\rm k}$, is derived, which
gives the disk aspect ratio
\begin{equation}
\label{eq:aspect}
h/r = 0.047 \left(\frac{r}{1~{\rm AU}}\right)^{1/4}
\left(\frac{L_*}{L_\odot}\right)^{1/8}
\left(\frac{M_*}{M_\odot}\right)^{-1/2},
\end{equation}
where $M_*$ and $M_\odot$ are the masses of the host star and the Sun, respectively.

In all of the
simulations, a Jovian--mass planet is initially placed at 0.05 AU, therefore
it is an HJ that can open an annular gap  around its orbit. 
\citet{crida_etal06} have derived the gap opening criterion for a viscous disk as
\begin{equation}
\frac{3}{4} \frac{h}{r_{\rm H}} + \frac{50}{q \mathcal{R}} \lesssim 1,
\end{equation}
where $r_{\rm H}=(q/3)^{1/3}a$ and $q\equiv M/M_*$ 
are the Hill radius of the planet with semimajor axis $a$ and the mass ratio of the planet to the star,
respectively. The Reynolds number is defined as 
$\mathcal{R} \equiv r^2 \Omega/\nu$, where the turbulent
viscosity prescription $\nu \equiv \alpha c_{\rm s}h $ is applied with $\alpha$
being a coefficient indicating the strength of turbulence.
The Jovian--mass planet at 0.05 AU satisfies this condition for almost any value
of $\alpha$.

\begin{figure}[htbp]
\epsscale{0.8}
\plotone{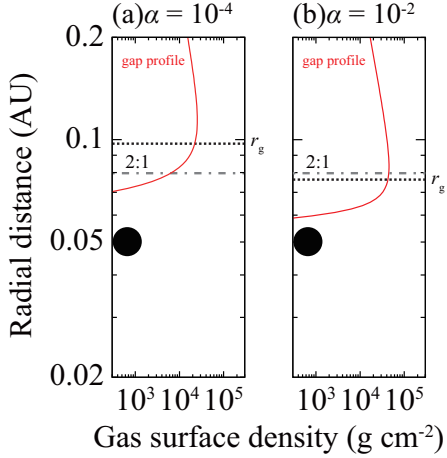}
\caption{Gas surface density profile for (a) $\alpha = 10^{-4}$ and (b) $\alpha = 10^{-2}$,
assuming $f_{\rm g}=1$. The dotted and dot--dashed lines indicate the locations of the gap
edge $(r_{\rm g})$ and the 2:1 MMR with the HJ at 0.05 AU, respectively.}
\label{fig:model}
\end{figure}

An analytical description for the computation of the gap
profile is also derived in Equation~(14) of \cite{crida_etal06}, in which the gradient of
gas density is given depending on the mass ratio and the disk properties.
Numerically
integrating this equation yields the gap profile of the gas surface density.
Figure~\ref{fig:model} shows the disk surface density profile 
for (a) the case of $\alpha = 10^{-4}$ and (b) the case of
$\alpha = 10^{-2}$.

The remarkable point in our simulations is the initial existence of an HJ 
in the inner cavity of a protoplanetary disk prior to the formation of rocky planets. 
Our result, however, does not depend on the origin of the HJ. Here we simply mention that, 
at least, the gravitational fragmentation of massive circumstellar disks and halfway migration 
of the resultant gaseous objects may provide a possible mechanism for providing the setup 
of our simulations.
Note also that the HJ at 0.05 AU may migrate further inward due to a one--sided
torque from the outer disk. However, the existence of numerous HJs may suggest that 
further migration seems limited.
In order to evaluate the torque, high--resolution 3D hydrodynamical simulations is required.
With the present computing power, it is not easy to resolve the gas flow around the HJ
that opens up a density gap around its orbit and to accurately derive the torque onto the HJ.
In this article, we assume for simplicity that the HJ is initially located at 0.05 AU.

\subsection{\textit{N}--body Code}
The orbits of embryos/planetesimals with masses $M_{1}, M_{2}, ...$ and position vectors
$\textbf{\textit{r}}_1, \textbf{\textit{r}}_2, ...$ relative to the host star 
are calculated by numerically integrating the equation of motion:
\begin{eqnarray}
\label{eq:eom}
\frac{d^2 \textbf{\textit{r}}_k}{dt^2}
& = & -GM_* \frac{\textbf{\textit{r}}_k}{ |\textbf{\textit{r}}_k|^3} 
- \sum_{j \neq k} GM_j 
\frac{\textbf{\textit{r}}_k - \textbf{\textit{r}}_j}{|\textbf{\textit{r}}_k - \textbf{\textit{r}}_j|^3} 
\nonumber\\
&   &
- \sum_{j} GM_j \frac{\textbf{\textit{r}}_j}{ |\textbf{\textit{r}}_j|^3} 
\nonumber\\
&   &
+ \textbf{\textit{F}}_{\rm damp} + \textbf{\textit{F}}_{\rm mig} + \textbf{\textit{F}}_{\rm aero}
+ \textbf{\textit{F}}_{\rm tide},
\end{eqnarray}
where $k, j$ = 1, 2, ..., the first term on the right--hand side 
is the gravitational force of the central star, the second term
is the mutual gravity between bodies, and the third is an indirect term. 
$\textbf{\textit{F}}_{\rm damp}$, $\textbf{\textit{F}}_{\rm mig}$, 
$\textbf{\textit{F}}_{\rm aero}$, and $\textbf{\textit{F}}_{\rm tide}$ represent specific forces for 
eccentricity damping, semimajor axis damping (type I migration) due to gravitational 
interaction with the disk gas, the aerodynamical gas drag force, and the tidal torque from the
central star, respectively 
(see \citealt{ogihara_ida09, ogihara_ida12}, and \citealt{ogihara_etal10} for each force formula).

Planetary embryos with masses $M$ perturb the disk gas and excite density waves, which
damp the orbital eccentricities, $e$, inclinations, $i$, and semimajor axes, $a$, of the embryos (e.g., \citealt{goldreich80}; \citealt{ward86}; \citealt{artymowicz93}). 
We use the formulation of \citet{tanaka_ward04} to calculate
$e$--damping and $i$--damping rates, and that of \citet{tanaka_etal02} for the
$a$--damping rate. 
The formulae for the specific tangential forces are given by
\begin{eqnarray}
\textit{F}_{{\rm damp}, r} &=& \frac{1}{0.78 t_e} \left(2A ^{c}_{r}\left[v_{\theta} - \sqrt{\frac{GM_*}{r}}\right] + A ^{s}_{r} v_r\right),\\
\textit{F}_{{\rm damp}, \theta} &=& \frac{1}{0.78 t_e} \left(2A ^{c}_{\theta}\left[v_{\theta} - \sqrt{\frac{GM_*}{r}}\right] + A ^{s}_{\theta} v_r\right),\\
\textit{F}_{{\rm damp}, z} &=& \frac{1}{0.78 t_e} \left(A ^{c}_{z}v_{z} + A ^{s}_{z} z \sqrt{\frac{GM_*}{r^3}} \right),\\
\textit{F}_{{\rm mig}, \theta} &=& \frac{1}{2 t_a} \sqrt{\frac{GM_*}{r}},
\end{eqnarray}
and $\textit{F}_{{\rm mig}, r} = \textit{F}_{{\rm mig}, z} = 0$,
where $v_r$, $ v_{\theta}$, and $v_{z}$ are the radial, tangential, and vertical
components of the planet's velocity.  
The numerical factors are given by $A^{c}_{r} = 0.057$, $A^{s}_{r} = 0.176$,
$A^{c}_{\theta} = -0.868$, $A^{s}_{\theta} = 0.325$, $A^{c}_{z} = -1.088$, and
$A^{s}_{z} = -0.871$ \citep{tanaka_ward04}, and $t_{e}$ and $t_{a}$ are given by
\begin{eqnarray}
t_e & = & \frac{1}{0.78}\left(\frac{M}{M_*}\right)^{-1} 
\left(\frac{\Sigma_{\rm g} r^2}{M_*}\right)^{-1}
\left(\frac{c_{\rm s}}{v_{\rm K}}\right)^{4} \Omega^{-1},\\
&=& 3 f_{\rm g}^{-1}
\left(\frac{r}{0.1~{\rm AU}}\right)^2
\left(\frac{M}{M_\oplus}\right)^{-1}
\left(\frac{M_*}{M_\odot}\right)^{-1/2}
\nonumber\\
&   &
\times \left(\frac{L_*}{L_\odot}\right)^{1/2}
{\rm ~yr},
\label{eq:e-damp}
\end{eqnarray}
and
\begin{eqnarray}
t_a & = & \frac{1}{C_{\rm I}}
\frac{1}{2.7+1.1q(r)} \left(\frac{M}{M_*}\right)^{-1}
\left(\frac{\Sigma_{\rm g} r^2}{M_*}\right)^{-1}
\left(\frac{c_{\rm s}}{v_{\rm K}}\right)^{2} \Omega^{-1},\\
&=& 1.6 \times 10^3 C_{\rm I}^{-1} f_{\rm g}^{-1}
\left(\frac{2.7+1.1q(r)}{4.35}\right)^{-1}
\left(\frac{r}{0.1~{\rm AU}}\right)^{3/2}
\nonumber\\
&   &
\times \left(\frac{M}{M_\oplus}\right)^{-1}
\left(\frac{M_*}{M_\odot}\right)^{-1/2}
\left(\frac{L_*}{L_\odot}\right)^{1/4}
{\rm ~yr},
\label{eq:a-damp}
\end{eqnarray}
where $v_{\rm K}$ is the Keplerian velocity and $\Omega$ is the Keplerian frequency.
Here, $-q(r)$ denotes the local surface density gradient 
($q(r) = -d \ln \Sigma_{\rm g}/d \ln r$).
We use the local value of $q(r)$ for each particle, thus $q(r) = 3/2$ when the bodies
are located away from the gap such that the planets migrate inward. 
In the vicinity of the gap, $q(r)$ becomes smaller
than $-2.7/1.1$ and the direction of migration can be outward. 
The location of zero migration at $r_{\rm g}$, $q(r_{\rm g})=-2.7/1.1$, is shown by the dotted 
line in Figure~\ref{fig:model}.

We introduce a scaling factor $C_{\rm I}$ that allows for the retardation and acceleration of type I migration.
Many simulations (e.g., \citealt{paardekooper_papaloizou09}; \citealt{paardekooper_etal10};
\citealt{masset_casoli10}; \citealt{baruteau_etal11})
have been carried out to determine the type I migration rate under
various disk conditions, and have claimed that migration can be much slower,
or even reversed, compared to the estimate by \citet{tanaka_etal02}. 
\citet{paardekooper_etal11} have found that when the timescales for viscous and
thermal diffusion inside the horseshoe region are comparable to the dynamical 
timescale, nonlinear effects dominate the corotation torque, leading to outward
migration. Although several authors add nonlinear correction factors to the migration 
timescale (e.g., \citealt{lyra_etal10}; \citealt{horn_etal12}; \citealt{hellary_nelson12}), 
for simplicity we instead adopt $C_{\rm I}$ as a parameter and examine the 
dependence of the migration rate on the final configuration of planets.

The force of aerodynamical gas drag acting on a body with mass $M$ and radius $R$
is given by $\textbf{\textit{F}}_{\rm aero} = -(1/2) C_{\rm D} \pi R^{2} \rho_{\rm gas}
|\textbf{\textit{v}} - \textbf{\textit{v}}_{\rm gas}| (\textbf{\textit{v}} - \textbf{\textit{v}}_{\rm gas})/M$
\citep{adachi_etal76}, where $C_{D} = 0.5$, $\rho_{\rm gas}$ is the gas density, 
$\textbf{\textit{v}}$ is the orbital velocity of the body, and $\textbf{\textit{v}}_{\rm gas}$ is the gas velocity.
The gas velocity is slightly different from the Keplerian velocity, $\textbf{\textit{v}}_{\rm K}$,
due to pressure gradientS;
$\textbf{\textit{v}}_{\rm gas} = (1-\eta)\textbf{\textit{v}}_{\rm K}$, where
\begin{eqnarray}
\eta = -\frac{1}{2} \left(\frac{c_{\rm s}}{v_{\rm K}}\right)^2
\frac{d \ln{P}}{d \ln{r}}.
\end{eqnarray}
Because $c_{\rm s} \propto r^{-1/4}$ from Equation~(\ref{eq:sound_speed}), $d \ln{P}/d \ln{r} = q(r) + 7/4$.
The aerodynamical drag becomes less effective than the gravitational
drag when $M \gtrsim 10^{-2}~M_\oplus$ at 
$a \simeq 0.1~{\rm AU}$ (e.g., \citealt{kominami_etal05}; \citealt{ogihara_ida09}), thus in the late stage of
planetary accretion, orbital evolution is mainly controlled by 
$\textbf{\textit{F}}_{\rm damp}$ and $\textbf{\textit{F}}_{\rm mig}$.

For numerical integration, we use a fourth--order Hermite scheme
\citep{makino_aarseth92} with a hierarchical individual time 
step \citep{makino91}.
When the physical radii of two spherical bodies overlap, 
they are assumed to merge into one body, conserving total mass and 
momentum assuming perfect accretion.
The physical radius of a body is determined by its mass, $M$,
and internal density, $\rho$, as
\begin{equation}
R = \left(\frac{3}{4 \pi}
\frac{M}{\rho}
\right)^{1/3},
\end{equation}
where we adopt $\rho = 3~{\rm g~cm}^{-3}$.
In large--number simulations, the radii of bodies are enhanced by a
factor of five (e.g., \citealt{kokubo_ida96}) to save computational time.

\subsection{Initial Conditions}
There are two types of models used for \textit{N}--body simulations in this paper, namely, 
simulations starting from
planetesimals (large--$N$ simulations) and simulations starting from protoplanets
expected to be formed from planetesimals (small--$N$ simulations).
In the former, we handle 5000 bodies in a single calculation,
which enables us to make detailed discussions of planetary accretion. However,
this calculation has a huge computational cost; a typical run uses about two--three months of CPU time on
special purpose machines for \textit{N}--body simulations (GRAPE--DR).
On the other hand, in the latter model $(N \sim 10)$, we are able to save
computational time, which allows the parameter space to be explored
much more efficiently, although this is not appropriate for investigating the
growth mode of planetesimals (e.g., oligarchic growth).
In this article, we first perform calculations starting from planetesimals for a fiducial case to examine
planetary accretion, and then focus on exploring parameter space using calculations
starting from protoplanets.

In this subsection, we describe the initial conditions for each method.
In both methods, the initial solid surface density is assumed to be
\begin{eqnarray}
\Sigma_{\rm d} =10 f_{\rm d} \left(\frac{r}{1~{\rm AU}}\right)^{-3/2}  {\rm ~g~cm^{-2}},
\label{eq:surface_solid}
\end{eqnarray}
where $f_{\rm d}$ is a scaling factor for the initial solid surface density. 
In the case of solar metallicity, $f_{\rm d} = f_{\rm g}$.
The initial conditions and values of input parameters for each run are summarized in 
Table~\ref{tbl:parameters}.

\begin{deluxetable}{lcccccc}
\tabletypesize{\footnotesize}
\tablecolumns{7}
\tablewidth{0pc}
\tablecaption{Simulation parameters}
\startdata
\hline \hline
Run		& $N_{\rm ini}$		& $C_{\rm I}$	& $f_{\rm d}$	& $f_{\rm g}$	& $t_{\rm dep}$ (yr)	&  $\alpha$\\
\hline
Aa1-3	& 5000	& 1	& 1	& 1	& $10^6$	& $10^{-4}$\\
Ba1-5	& 20		& 1	& 1	& 1	& $10^6$	&  $10^{-4}$\\
Bb1-5	& 20		& 0.1	& 1	& 1	& $10^6$	& $10^{-4}$\\
Bc1-5	& 20		& 10	& 1	& 1	& $10^6$	& $10^{-4}$\\
Bd1-5	& 40		& 1	& 0.1	& 1	& $10^6$	& $10^{-4}$\\
Be1-5	& 7		& 1	& 10	& 1	& $10^6$	& $10^{-4}$\\
Ca1-3	& 20		& 1	& 1	& 0.01	& $10^6$	& $10^{-4}$\\
Cb1-3	& 20		& 1	& 1	& 0.1	& $10^6$	& $10^{-4}$\\
Cc1-3	& 20		& 1	& 1	& 10		& $10^6$	& $10^{-4}$\\
Da1-3	& 20		& 1	& 1	& 1	& $3\times10^6$	& $10^{-4}$\\
Db1-3	& 20		& 0.1	& 1	& 1	& $3\times10^6$	& $10^{-4}$\\
Dc1-3	& 20		& 10	& 1	& 1	& $3\times10^6$	& $10^{-4}$\\
Dd1-3	& 40		& 1	& 0.1	& 1	& $3\times10^6$	& $10^{-4}$\\
De1-3	& 7		& 1	& 10	& 1	& $3\times10^6$	& $10^{-4}$\\
Ea1-3	& 20		& 1	& 1	& 1	& $10^6$	& $10^{-2}$\\
Eb1-3	& 20		& 0.1	& 1	& 1	& $10^6$	& $10^{-2}$\\
Ec1-3	& 20		& 10	& 1	& 1	& $10^6$	& $10^{-2}$\\
Ed1-3	& 40		& 1	& 0.1	& 1	& $10^6$	& $10^{-2}$\\
Ee1-3	& 7		& 1	& 10	& 1	& $10^6$	& $10^{-2}$
\enddata
\tablecomments{List of parameters for each simulation: the number of bodies that are
initially placed between 0.1 and 0.5 AU, $N_{\rm ini}$; the type I migration efficiency factor,
$C_{\rm I}$; the scaling factor for the solid surface density, $f_{\rm d}$; the scaling factor for
the gas surface density, $f_{\rm g}$; the disk depletion timescale, $t_{\rm dep}$; and
the scaling factor for the disk viscosity, $\alpha$. Three runs are performed
for each model, except for runs Ba, Bb, Bc, Bd and Be, where five runs are carried out.
Our fiducial runs are Aa1--3 and Ba1--5.}
\label{tbl:parameters}
\end{deluxetable}

\subsubsection{Simulations Starting from Planetesimals}
Initially, 5000 planetesimals with mass $M = 2 \times 10^{24} {\rm ~g}$ are placed between
$a = 0.1-0.5 {\rm ~AU}$.  The magnitude of the initial velocity dispersion is equal to the
escape speed of those planetesimals.

\subsubsection{Simulations Starting from Protoplanets}
For small--$N$ calculations, our initial conditions start 
with planetary embryos that arise from oligarchic growth \citep{kokubo_ida98},
which is the same as in OIK13.
If planetary embryos fully accrete the surrounding planetesimals, they eventually have
an isolation mass given by
$M_{\rm iso} = 2 \pi a \Delta a \Sigma_{\rm d},$ 
where $\Delta a$ is the width of an embryo's feeding zone. Using 
Equation~(\ref{eq:surface_solid}) for $\Sigma_{\rm d}$,
\begin{eqnarray}
M_{\rm iso} &=&  0.16 f_{\rm d}^{3/2} \left(\frac{a}{1 {\rm ~AU}}\right)^{3/4}
\left(\frac{\Delta a}{10 r_{\rm H}}\right)^{3/2}
\nonumber \\&& \times
\left(\frac{M_*}{M_\odot}\right)^{-1/2}
M_\oplus.
\label{eq:m_iso}
\end{eqnarray}
However, an embryo may start migration before
it reaches the isolation mass. The critical mass for migration is derived from
the balance between the migration timescale, $t_a$, and the accretion timescale, $t_{\rm acc};$
$t_{\rm acc} = t_a$.
Here, $t_{\rm acc} \simeq (\Sigma_{\rm d} \Gamma v_{\rm ran} /h_{\rm d} M)^{-1}$ \citep{safronov69}, 
where $\Gamma$
is the cross section between embryos and planetesimals, $v_{\rm ran}$ is
the velocity dispersion, and $h_{\rm d}$ is the scale height of the
planetesimal disk.  The velocity dispersion is determined by the equilibrium between
viscous stirring by embryos and damping by gas drag, gravitational focusing is taken into
account for $\Gamma$, $h_{\rm d} \simeq a(v_{\rm ran}/v_{\rm K}),$ and thus the accretion timescale is given by
\citep{kokubo_ida02}:
\begin{eqnarray}
t_{\rm acc} &=& 1.4 \times 10^5 f_{\rm d}^{-1} f_{\rm g}^{-2/5}  
\left(\frac{a}{1 {\rm ~AU}}\right)^{27/10}
\left(\frac{M}{M_\oplus}\right)^{1/3}
\nonumber\\
&   & \times
\left(\frac{M_*}{M_\odot}\right)^{-1/6}
\left(\frac{\rho}{3 {\rm~g~cm^3}}\right)^{1/3}
\left(\frac{m}{10^{18} {\rm ~g}}\right)^{2/15}
{\rm yr},
\label{eq:t_acc}
\end{eqnarray}
where $m$ is the mass of planetesimals.
Thus the critical mass is 
\begin{eqnarray}
M_{\rm crit} &=&  0.47 f_{\rm d}^{3/4} f_{\rm g}^{-9/20} C_{\rm I}^{-3/4} 
\left(\frac{a}{1 {\rm ~AU}}\right)^{-9/10}
\left(\frac{M_*}{M_\odot}\right)^{-1/4}
\nonumber\\
&   & \times
\left(\frac{\rho}{3 {\rm~g~cm^3}}\right)^{-1/4}
\left(\frac{m}{10^{18} {\rm ~g}}\right)^{-1/10}
M_\oplus.
\label{eq:m_crit}
\end{eqnarray}
We use  the smaller of $M_{\rm iso}$ and $M_{\rm crit}$
for the mass of initial embryos. As stated in OIK13,
$M_{\rm iso}$ is smaller than $M_{\rm crit}$ in the inner region, therefore we usually start
with isolation--mass embryos. 
As seen below in the results of \textit{N}--body simulations, we find that 
the resultant orbital configuration of formed planets does not depend on the 
initial mass of planetesimals.
We set the initial
eccentricities and inclinations of embryos to be as small as $10^{-2}$ with the relation
$\langle e^{2} \rangle ^{1/2} = 2 \langle i^{2} \rangle ^{1/2}$.

\subsection{Hybrid Scheme Combining \textit{N}--body Code with Semianalytical Evolution}
We set the calculation region for our $N$--body simulations between 0.02--0.5 AU because we 
focus on the formation of close--in terrestrial planets.
Meanwhile, because of the effect of type I
inward migration, embryos formed in the disk beyond 0.5 AU invade
the calculation domain as long as embryos migrate due to interactions with 
the gas disk.
It would be worthwhile to extend the calculation region beyond 0.5 AU
(say, to 2 AU), however, the significant computational cost
for calculating mutual gravitational interactions makes this impossible.
Hence, previous \textit{N}--body simulations only followed the orbits of 
the initially placed bodies (e.g.,\citealt{thommes05};
\citealt{terquem_papaloizou07};  \citealt{ogihara_ida09}).

In our simulations, the evolution of embryos beyond 0.5 AU
is calculated not by using the
gravitational \textit{N}--body code but by semianalytically simulating the growth
and migration of solid bodies.
That is, the gravitational \textit{N}--body simulation and the semianalytical 
simulation are performed simultaneously; when embryos that are calculated
using the semianalytical code reach the boundary ($a = 0.5 {\rm AU}$) they are then 
added to the \textit{N}--body code.

In the outer disk beyond 0.5 AU,
the growth and migration of embryos are calculated in the same way
as the population synthesis models (e.g., \citealt{ida_lin04}; \citealt{ida_lin08}).
The growth of planetary embryos occurs within the timescale given in
Equation~(\ref{eq:t_acc}) and terminates when they reach the isolation 
mass (Equation~(\ref{eq:m_iso})). These embryos migrate within the timescale
of Equation~(\ref{eq:a-damp}).

\section{RESULTS: SIMULATIONS STARTING FROM PLANETESIMALS}
\label{sec:results1}
\begin{figure}[htbp]
\epsscale{1.0}
\plotone{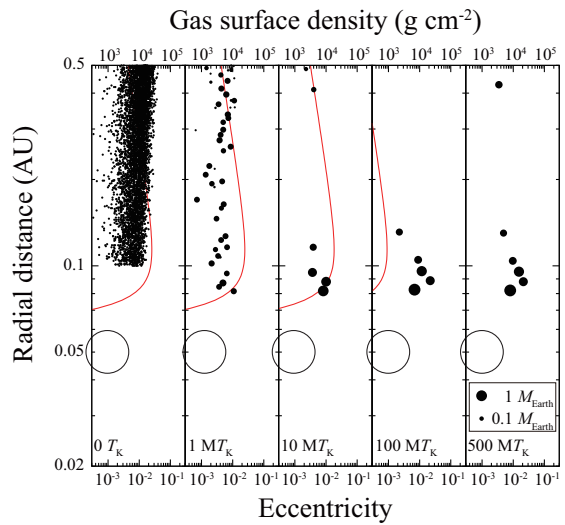}
\caption{
Snapshots of a system for run Aa1. Filled circles represent bodies and open circles show
the HJ. The solid line indicates the gas surface density (upper axis).}
\label{fig:snap}
\end{figure}

\begin{figure}[htbp]
\epsscale{0.9}
\plotone{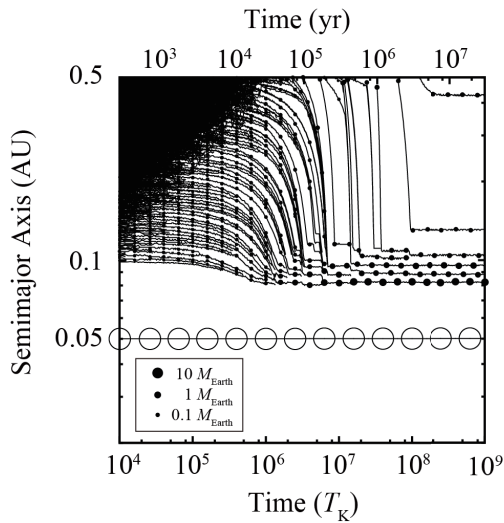}
\caption{Time evolution of planets for run Aa1. 
The filled circles connected with solid lines represent the bodies, while the open
circles show the HJ.}
\label{fig:t_am}
\end{figure}

We first present the results of high--resolution \textit{N}--body simulations 
for a fiducial model with the goal of revealing the formation process and
final properties of planetary systems.
Figure~\ref{fig:snap} shows snapshots in time of the evolution of one
simulation (run name: Aa1). 
Figure~\ref{fig:t_am} shows the time evolution of the semimajor axis.

The formation of terrestrial planets from planetesimals under the influence of a
giant planet can be understood in three stages. In the first stage $(t/T_{\rm K} \lesssim 10^{5})$, 
planetesimals grow to planetary embryos or protoplanets via runaway/oligarchic growth. Planetesimals 
grow from the inside out because of the high solid surface density and short 
orbital period in the inner region. 
Protoplanets born in a swarm of planetesimals have low $e$ and $i$ through energy 
equipartition with planetesimals. As protoplanets grow, because planetesimals have high $e$ and $i$,
energy equipartition is lost. Then slow oligarchic growth starts
(\citealt{ida_makino93}; \citealt{ormel_etal10}).
Protoplanets keep their orbital separations at about $\simeq 8 r_{\rm H}$,
which is slightly smaller than that at 1 AU (e.g., \citealt{kokubo_ida98}).

The second stage $(10^{5} \lesssim t/T_{\rm K} \lesssim 10^{8})$ is the migration phase 
leading to close scattering and/or formation of MMRs between planets: 
orbital configurations are significantly altered and final
orbits are almost built up. Planetary embryos start migration when they reach
the isolation mass, $M_{\rm iso}$, or the critical mass for migration, $M_{\rm crit}$, whichever is smaller.
The isolation mass usually determines the masses of embryos starting migration
in the inner disk (see Equations~(\ref{eq:m_iso}) and (\ref{eq:m_crit}) and OIK13). 
The first--born innermost protoplanet reaches the edge of
the gap at $t \simeq 10^6~T_{\rm K}$ and ceases its migration due to positive torque 
from the disk.
Note that planets cannot migrate to the location of the 2:1 MMR
with the HJ because the position of the 2:1 resonance $(\simeq 0.08 {\rm AU})$ is inside the
gap edge in the fiducial model ($\alpha=10^{-4}$; Figure~\ref{fig:model}(a)).
The time of the onset of the migration phase increases with $a$, and outer protoplanets
sequentially migrate inward even after inner protoplanets cease migration at the gap edge. 
When migrating protoplanets subsequently approach the inner planet
trapped at the gap edge, they experience a close encounter and merge into
one body, or are captured in mutual MMRs.
The protoplanet captured in the MMR also encounters a protoplanet newly approaching from the
outer disk resulting in a collision or a resonance capture.
The repetition of resonance captures establishes a chain 
of resonant planets.
The properties of the resonant chain (e.g.,
commensurate values) depend on the formation conditions, which are discussed
in Section~\ref{sec:resonance}. 
Even in a resonant chain, close scattering and collisional coagulation result in a relocation 
of protoplanets. Eventually, the largest planet is located at the innermost
orbit of the resonant chain.
This stage lasts until the disk gas decays enough that protoplanets no longer
move to the inner region $(t \simeq 10^8~T_{\rm K} \simeq 3 \times 10^6~{\rm yr})$. 

In the third stage $(t/T_{\rm K} \gtrsim 10^{8})$, the disk gas fully decays and in some 
cases planets exhibit
orbit crossing resulting in giant impacts between planets.
Since the damping force also vanishes due to the gas depletion, 
the eccentricities of planets can effectively increase through
mutual interactions, which enables collisions between planets (giant impacts).
Whether planets in a resonant chain
undergo orbit crossing depends on the properties of the system 
(e.g., orbital separation and number of planets). 
In run Aa1, giant impacts between planets
are not observed before $t = 5\times 10^8~T_{\rm K}$. 
We continued this 
simulation until $t = 10^9~T_{\rm K}$ but never saw a giant impact.
However, for other large--$N$ simulations (runs Aa2 and Aa3), orbit crossing and resultant giant impacts between
planets do occur. Note that even if damping due to gas drag is absent, 
planets do not
have high eccentricities; the maximum eccentricity is about 0.05, therefore
they do not exhibit ``global'' orbit crossing but only collide with their neighboring 
planets, which is also shown by \citet{ogihara_ida09}.
As a result, some commensurate relations between planets that have not experienced
giant impacts can remain at the end.

By the end of the simulations, six planets, not including the HJ, have formed
inside 0.5 AU. 
The inner four planets can be captured in mutual MMRs to
make a resonant chain:
the innermost and second innermost planets have 10:9 commensurability, 
the second and third planets have 9:8 commensurability, and 
the third and fourth planets have 8:7 commensurability. 
Since these are relatively closely spaced resonances, resonant angles are
not always librating around some fixed values but circulating.
The eccentricities are relatively small $(\simeq 0.01)$.
The total mass in planets excluding the HJ is
$3.2 M_\oplus$, of which $1.5 M_\oplus$ has migrated from outside 0.5 AU.
The largest planet has
a mass of $1.6 M_\oplus$ and is located at $0.082 {\rm AU}$.
This is slightly outside the location of the 2:1 MMR ($\simeq 0.08 {\rm AU}$) with
the HJ. 
The mass ratio between the largest mass and the total mass
is 0.49; up to 50 percent of solid materials in the close--in region is accumulated
by the largest planet, and there is an order--of--magnitude difference in mass 
among the formed planets. 

The final properties of each run independently
starting from different random initial planetesimal positions in the \textit{N}--body 
calculation $(a \leq 0.5~{\rm AU})$ and the semianalytical calculation $(a > 0.5~{\rm AU})$
are summarized in Table~\ref{tbl:results}.
Orbital configurations at $t = 5 \times 10^8 T_{\rm K}$ are also shown.
For the results with fiducial parameters (runs Aa1--3 and Ba1--5), the simulation is performed
until $t = 10^9 T_{\rm K}$, thus orbital properties at $t=10^9 T_{\rm K}$ are also presented
in Table~\ref{tbl:results}.
We find that several (three to six) planets formed that are in
MMRs making resonant chains. These commensurate values
were predicted by a previous study of capture into MMRs
\citep{ogihara_kobayashi13}, which is described in Section~\ref{sec:resonance}.
Note that in run Aa3, giant impact events 
occur during the gas depletion phase  $(t = 4.1 \times 10^8 T_{\rm K})$ and resonant
relations are lost.
Eccentricities are relatively small $(\simeq 0.01)$ even after the $e$-damping
force due to the gas disk vanishes. The largest planets, which are located at
$\simeq 0.08-0.09 {\rm AU}$, have mass $\simeq 2.3 M_\oplus$ constituting 
$\ga 50$ percent of the total mass inside 0.5 AU.

\section{RESULTS: SIMULATIONS STARTING FROM PROTOPLANETS}
\label{sec:results2}
We next present the results of our simulations that reduce the number of simulated
bodies and vary the model parameters over a wide range. 
See Table~\ref{tbl:parameters} for details of the parameters.

\subsection{Fiducial Model}
\begin{figure}[htbp]
\epsscale{0.9}
\plotone{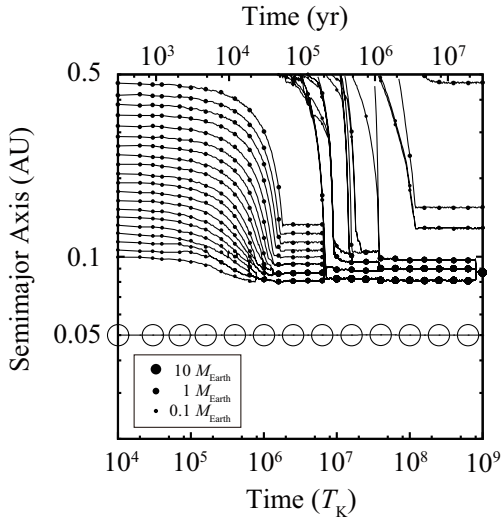}
\caption{Result of the run Ba1 simulation (fiducial model).}
\label{fig:t_am_fid}
\end{figure}

First, simulations for a fiducial model (runs Ba1--5) are performed, in which we adopt 
the same values for parameters $C_{\rm I}, f_{\rm g}, f_{\rm d},
t_{\rm dep},$ and  $\alpha$ (but not $N_{\rm ini}$) as in runs Aa1--3.
Figure~\ref{fig:t_am_fid} shows the evolution of the semimajor axis
for run Ba1 for comparison with the large--$N_{\rm ini}$ result shown in 
Figure~\ref{fig:t_am}.

As for the formation
process, the first embryo formation stage is not calculated in this run. 
Therefore protoplanets at around 0.5 AU migrate earlier than those in large--$N$
simulations. In addition, the masses of migrating protoplanets 
that are initially placed inside 0.5 AU are slightly 
larger than those in large--\textit{N} simulations (runs Aa1--3). Since we
adopt relatively large orbital separations between protoplanets 
($\Delta a \simeq 15 r_{\rm H}$) as the initial condition to reduce the computational
cost, the isolation mass becomes slightly larger. Although the differences in
mass of the migrating bodies are within a factor of two, planet migration is accelerated. 
This early planetary migration makes an unphysical time--gap in
Figure~\ref{fig:t_am_fid}(a) at around $t \sim 10^{6} T_{\rm K}$, between
the migration phase of bodies that initially reside within 0.5 AU and those migrating from
outside 0.5 AU. Note, however, that this produces no systematic change in the final state of
planetary systems. In fact, after the emergence of protoplanets from the outer region
$(t \ga 10^7~T_{\rm K})$, the evolution is the same: migrating bodies interact with
the inner planets in the resonant chain, which are trapped by the gap edge or captured in MMRs, 
leading to rearrangements of the resonant chain. By the time no more embryos
are migrating inward and the migration stage is over $(t \simeq 10^8 T_{\rm K})$, 
several planets are lined up by the gap edge captured in the resonant chain.

In the final stage, when the gas disk is depleted, there is no orbit crossing and hence no giant
impacts occur in run Ba1 before $t = 5 \times 10^8~T_{\rm K}$. 
In run Ba4, however, local orbit
crossings and giant impacts are observed at $t \simeq 2.7 \times 10^8~T_{\rm K}$.
Although some MMRs are destroyed through this orbital instability,
one resonant relation (a 4:3 resonance between the innermost planet and the second
innermost planet) remains.
Note that we continue the calculation until $t = 10^9 T_{\rm K}$ and observe
giant impacts after $t = 5 \times 10^8~T_{\rm K}$
in several runs (Ba1, Ba2 and Ba5). In these cases, some resonant configurations
can be destroyed. 

The final state for each run, starting from different positions,
is summarized in Table~\ref{tbl:results}, in which there is no
significant difference between runs Aa1--3 and Ba1--5. 
In every simulation, several planets (three to eight) form that are partially
captured in MMRs. The typical value for the resonant commensurability is about 7:6,
although the resonant angles are not necessarily 
librating. 
Note that several planets can be captured into coorbital resonances with each other
during the migration phase, which has been shown in previous studies
(\citealt{thommes05}; \citealt{jakubik_etal12}). However, such 1:1 resonances are lost
by the end through gravitational perturbations from other migrating planets.
We observe several collision events during the gas dissipation phase, however, 
planets do not exhibit global orbital instability so some resonant relations tend to 
remain at the end. The final eccentricities of planets are small $(\simeq 0.01)$.
The mass of the largest planet is about $1.9~M_\oplus$ and consists of 40--80 percent
of the total mass in the close--in region. Thus, the mass ratio between the largest planet and
the other small planets is about a factor of 10. Such properties of the resulting planets are
mainly determined during the migration and final stages.
Therefore, the final masses and configurations of planets are almost independent of the initial
number of planetesimals, as shown in Table~\ref{tbl:results}.

\subsection{Dependence on Migration Efficiency}

\begin{figure*}[htbp]
\epsscale{0.8}
\plottwo{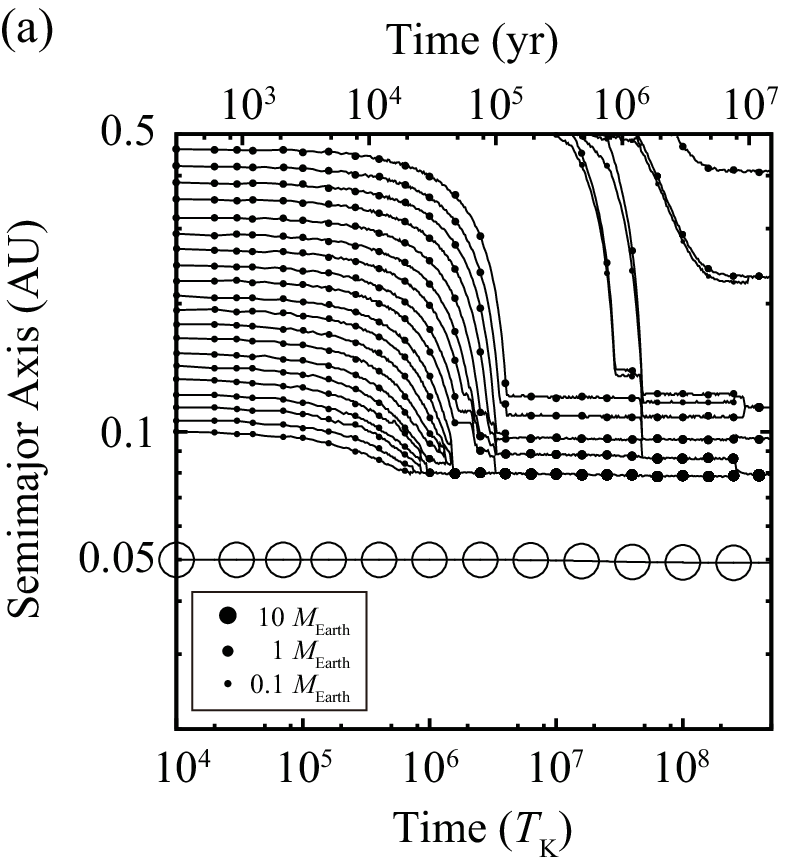}{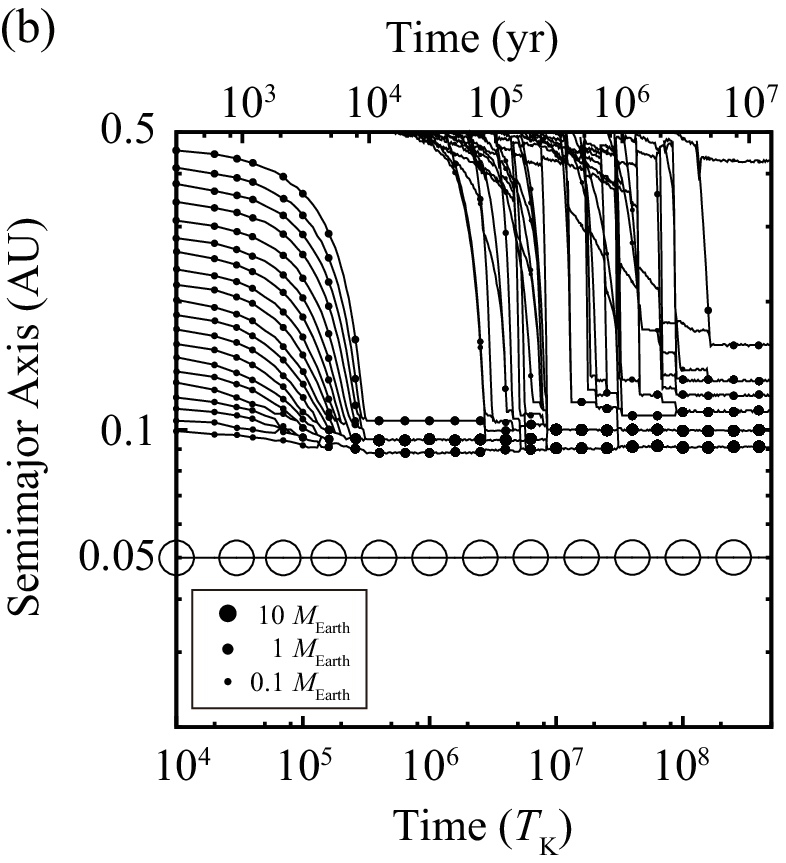}
\caption{Results of runs (a) Bb1 $(C_{\rm I}=0.1)$ and (b) Bc1 $(C_{\rm I}=10)$ simulations,
in which the efficiency of type I migration is changed.}
\label{fig:t_a_mig}
\end{figure*}

Figure~\ref{fig:t_a_mig} show the evolution of the semimajor
axis, where figures marked with (a) and (b) are the cases of $C_{\rm I} = 0.1$
(run Bb1; migration is less efficient) and $C_{\rm I}=10$ (run Bc1; migration is efficient), 
respectively. The initial masses of the 
embryos inside 0.5 AU are the same for the two cases.
Although the speed difference of type I migration is a 
factor of 100 between the two cases, the actual speed difference is less than a factor of 10. 
This is because for $C_{\rm I}=0.1,$ the migration caused by $\textbf{\textit{F}}_{\rm mig} (\propto C_{\rm I})$
is slower than that by $\textbf{\textit{F}}_{\rm damp}$.

If $e \neq 0$ or $i \neq 0$,
$\textbf{\textit{F}}_{\rm damp}$ is exerted and mainly acts to damp $e$ and $i$. 
We note that the torque caused by $\textit{F}_{{\rm damp},\theta}$ results in migration.
Taking the orbital average of torques, 
$r \textit{F}_{{\rm mig},\theta} < r \textit{F}_{{\rm damp},\theta}$ can be written as 
$C_{\rm I} \lesssim -4A^{c}_{\theta} e^{2} (c_{\rm s}/v_{\rm K})^{-2}/(2.7 + 1.1 q)
\simeq 2300 e^{2} (r/0.1 {\rm AU})^{-1/2}$. In the migration stage, embryos have eccentricities
of 0.01 (see Figure~\ref{fig:snap}). Therefore, when $C_{\rm I} \lesssim 0.1$, the migration 
caused by $F_{{\rm damp},\theta}$ is much more effective than that by 
$F_{{\rm mig},\theta}$.

In addition to this effect, we observe that the velocity components of bodies are altered due to
the existence of the HJ, which results in the loss of angular momentum through 
$\textbf{\textit{F}}_{\rm damp}$. The local velocity of the body at the apocenter is slower than the local
Keplerian velocity in the inertial frame $(= \sqrt{GM_* (1-e)/ a(1+e)})$ while the velocity of the body at 
the pericenter almost equals the local Keplerian velocity, thus on average the body gains a negative 
torque and undergoes inward migration. In the results for $C_{\rm I}=0.1$, this effect is
more effective.
We discuss this $e$--damping inducing migration in detail in Section~\ref{sec:velocity_difference}.

The planet formation process shown in Figure~\ref{fig:t_a_mig} is the same as for the fiducial model; before $t \simeq 10^8~T_{\rm K}$,
migrating planetary embryos interact with the planets in the resonant chain, in which the 
innermost planet is trapped by the gap edge. In the disk depletion stage, in some runs,
giant impacts between planets are observed. 
For both $C_{\rm I}=0.1$ and 10, four out of
five runs exhibit local orbit crossing and such collisions after $t = 10^8~T_{\rm K}$.
Even though some commensurabilities are lost
due to the mutual collisions, at least one resonant relation remains for each run.

The results of runs for $C_{\rm I}=0.1$ and 10 (Bb1--5 and Bc1--5) are summarized in Table~\ref{tbl:results}.
The final results are almost independent of the migration efficiency.
On average, seven planets are formed to make a resonant chain. The typical resonant 
values are between 4:3 and 8:7, which are almost identical among $C_{\rm I}=0.1$ and 10.
The difference in the migration speed between runs for $C_{\rm I}=0.1$ and 10 is within a factor of a few, which results in 
almost the same resonant configuration (e.g., commensurability). This will be discussed
in more detail in Section~\ref{sec:resonance}.

The final mass is also nearly the same among these runs. The total mass at the end of the $C_{\rm I}=10$ 
run is slightly larger than that for $C_{\rm I}=0.1$, however, the difference is only within a factor of 1.5 
even when the difference in $C_{\rm I}$ is a factor of 100. The same is true for the maximum mass.
The fraction of mass in the largest planet, $M_{\rm max}/M_{\rm tot}$, is about 0.5 on average, 
which is also comparable to the fiducial model.

\subsection{Dependence on Solid Surface Density}

\begin{figure*}[htbp]
\epsscale{0.8}
\plottwo{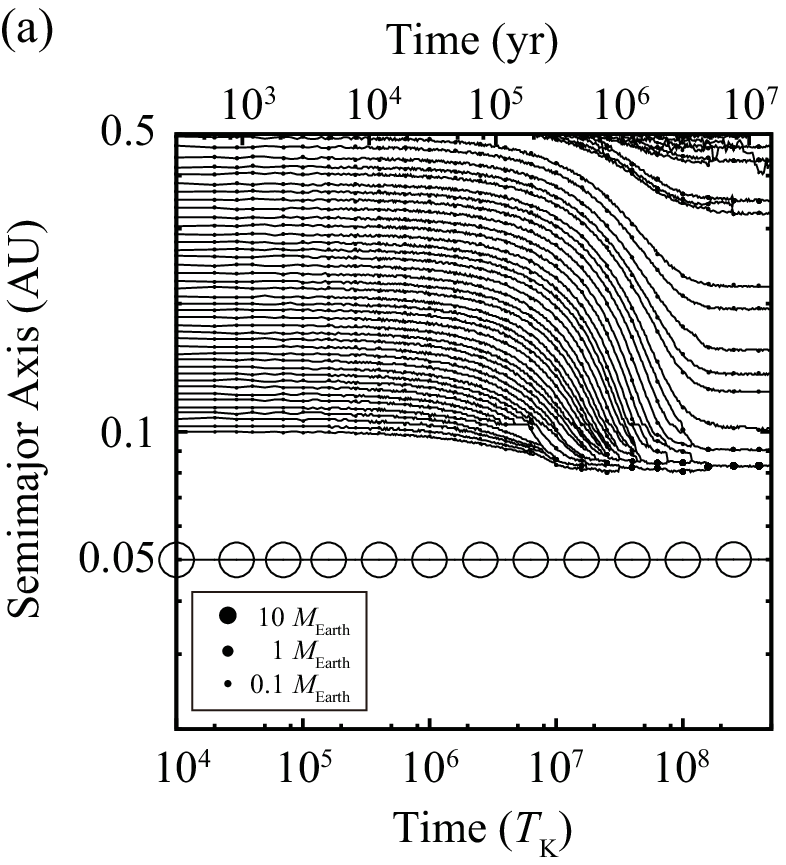}{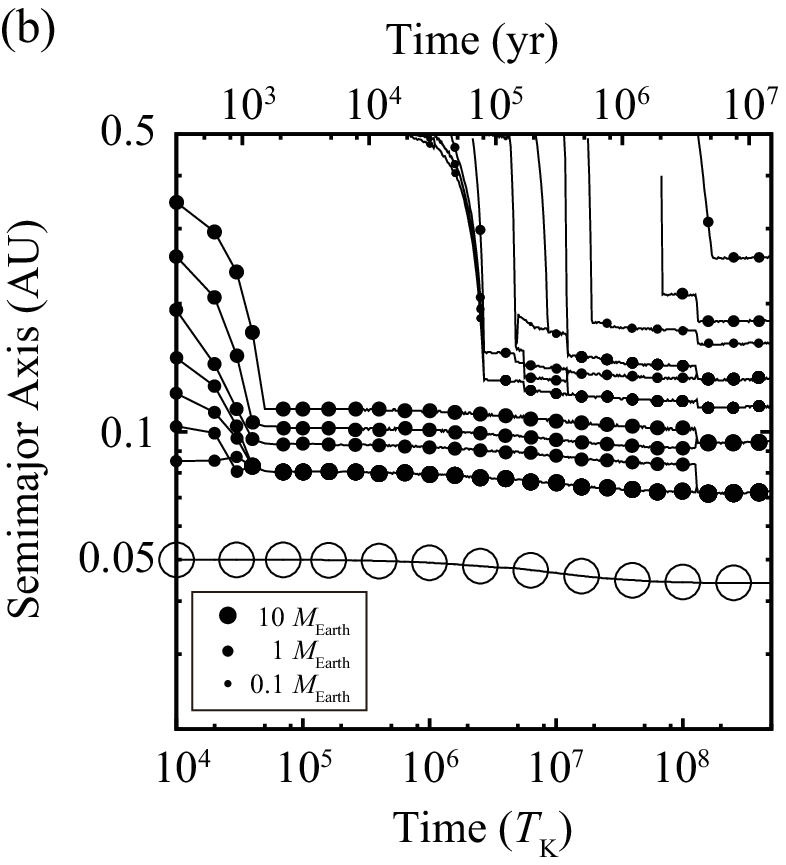}
\caption{Results of runs (a) Bd1 $(f_{\rm d}=0.1)$ and (b) Be1 $(f_{\rm d}=10)$ simulations,
in which the scaling factor for the initial solid surface density is changed.}
\label{fig:t_a_fd}
\end{figure*}

Next, we examine the dependence on solid surface density. Figure~\ref{fig:t_a_fd}
displays the evolution of $a$, and the panels
marked with (a) and (b) are the results for run Bd1 ($f_{\rm d}=0.1$) 
and run Be1 ($f_{\rm d}=10$), respectively.
There are differences in the initial distribution (e.g., number and mass) of planetary embryos 
inside 0.5 AU between the two cases.
The embryo masses differ by a factor of 1000 leading to a difference in migration
speed of the same order of magnitude. The amount of solid material also changes the growth
timescale of planets located beyond 0.5 AU. The total mass of solid bodies migrating from beyond 0.5 AU is
$\simeq 0.04 M_\oplus$ for $f_{\rm d}=0.1$ and $\simeq 5 M_\oplus$ for $f_{\rm d}=10$.
Although the planet formation process described above is not changed, in the results
for $f_{\rm d}=0.1$, small protoplanets are formed and,
hence, their migration timescales are so long that many planets
do not reach the vicinity of the gap edge before gas depletion. As a result,
orbital rearrangements of the resonant chain occur less often.
On the other hand, for the case of Be1 $(f_{\rm d}=10)$, the innermost planet is pushed inward by the
outer planets in the resonant chain and passes through the gap trap.
This is because the positive gap torque exerted on the innermost planet cannot compensate
for the total negative torque exerted on the other planets. 
In this case, the HJ undergoes slight inward migration by being pushed by
the chain of the resonant planets.
In the results for $f_{\rm d}=10$, several giant impacts are observed during the gas depletion stage,
with a higher frequency than in the fiducial model.
This can be attributed to large planets formed for $f_{\rm d}=10$:
The orbital separations scaled by their mutual Hill radii are small, which shortens
the orbital stability time (\citealt{chambers_etal96}; \citealt{matsumoto_etal12}).

The results of all runs (Bd1--5 and Be1--5) are summarized in Table~\ref{tbl:results}.
The number of final planets is between three and seven for $f_{\rm d}=10$,
which is comparable to that for the fiducial model. For the case of $f_{\rm d}=0.1$,
more planets tend to remain in the final state.
MMRs are seen in all runs except Be2, in which several giant
impact events occur at $t \simeq 3.8 \times 10^8~T_{\rm K}$ and resonant
relations are destroyed. The resonant values are also the same as those
shown above, but in the case of $f_{\rm d}=10$, planets are captured in
slightly separated resonances (e.g., 3:2 and 4:3).
There is a large difference in mass; the masses of the largest planets are
$\simeq 0.1 M_\oplus$ and $\simeq15 M_\oplus$ for the cases of $f_{\rm d}=0.1$ and 10,
respectively. The total masses are $\simeq 0.23 M_\oplus$ and $\simeq 25 M_\oplus$.
The maximum mass to total mass ratio is about 0.3--0.8.
The eccentricities of the planets increase with increasing $f_{\rm d}$:
Typical eccentricities for $f_{\rm d}=0.1$ are $\simeq 0.005$, whereas those for
$f_{\rm d}=10$ are $\simeq 0.03$.

\subsection{Gas Surface Density}
\label{sec:gas_density}

\begin{figure*}[htbp]
\epsscale{0.35}
\plotone{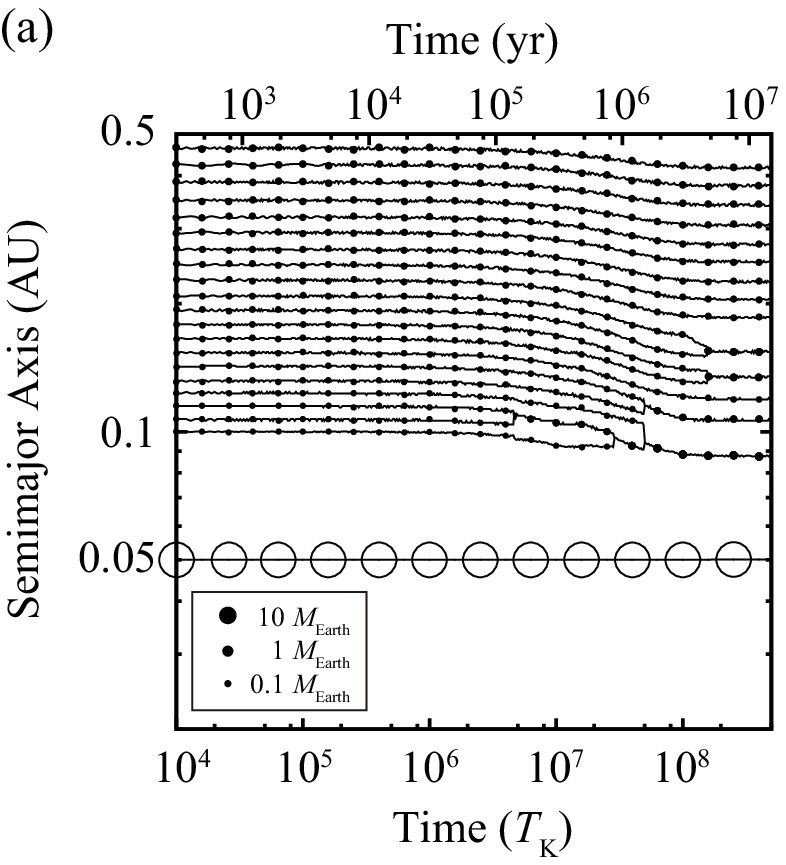}
\plotone{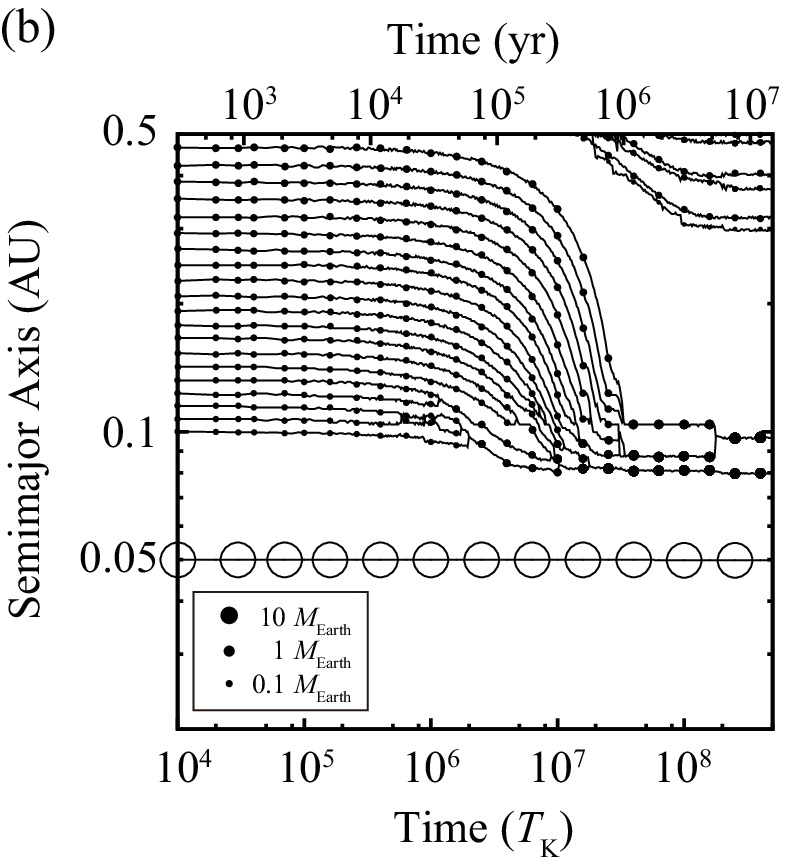}
\plotone{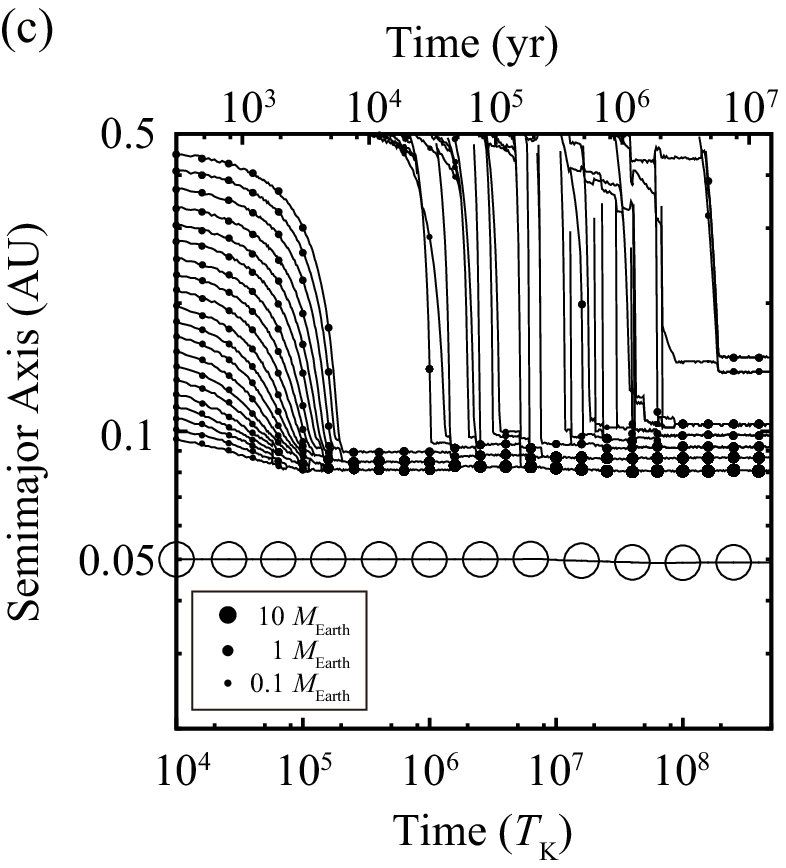}
\caption{Results of runs (a) Ca1 $(f_{\rm g}=0.01)$, (b) Cb1 $(f_{\rm g}=0.1)$, and (c) Cc1 $(f_{\rm g}=10)$ simulations, in which the scaling factor for the gas surface density is changed.}
\label{fig:t_a_fg}
\end{figure*}

Figures~\ref{fig:t_a_fg}(a), (b), and (c) display results for runs Ca1 $(f_{\rm g}=0.01)$, 
Cb1 $(f_{\rm g}=0.1)$, and Cc1 $(f_{\rm g}=10)$, respectively.
The decrease/increase in $f_{\rm g}$ corresponds to weakening/strengthening of both 
the type I migration and the eccentricity damping due to disk gas. 

For $f_{\rm g}=0.01$, the migration timescale becomes long and planets do 
not easily migrate inward, which decreases the total mass and the maximum mass.
The maximum mass and the total mass are typically $0.2~M_{\oplus}$ and 
$1.9~M_{\oplus}$, respectively.
The orbital rearrangement of the resonant chain during the migration phase is so ineffective that 
accumulation of solid materials at the gap edge does not occur, leading
to a smaller value of $M_{\rm max}/M_{\rm tot} \simeq 0.1$.
Several equal--mass planets, $N\simeq 13$, form and they tend to be in relatively separated 
MMRs; the typical value for the commensurability is 5:4.

However, the results for $f_{\rm g}=0.1$ and $10$ are basically similar to those of the fiducial model
$(f_{\rm g}=1)$. 
For large $f_{\rm g}$, migration effectively provides material at the edge of the gap and 
hence the masses of planets ($M_{\rm max}$ and $M_{\rm tot}$) increase with $f_{\rm g}$.

\subsection{Gas Dispersal Timescale}
We also perform simulations (Da1--3, Db1--3, Dc1--3, Dd1--3, De1--3) 
in which we adopt a gas dissipation timescale, $t_{\rm dep} = 3 \times 10^6 {\rm yr} \simeq 9 \times
10^7 T_{\rm K}$, longer by a factor of three
than that of the fiducial model (runs Ba1--5, Bb1--5, Bc1--5, Bd1--5, Be1--5 correspond
to simulations with shorter $t_{\rm dep}$).
In these simulations, we integrate the orbits of bodies until $t = 10^9 T_{\rm K}$.
The results of all runs are summarized in Table~\ref{tbl:results}.

The difference is the duration of the migration phase. For Da1, the total mass in planets that migrate
to the vicinity of the gap edge increases $(M_{\rm tot} \simeq 4.7 M_{\oplus})$, 
compared to short $t_{\rm dep}$ cases ($M_{\rm tot} \simeq 3.3 M_{\oplus}$ for run Ba1).
The mass of the largest planet at $t=10^9 T_{\rm K}$ is $M_{\rm max} \simeq 1.9 M_{\oplus}$,
which is comparable to or slightly larger than those for run Ba1--5.
This is because giant impacts between planets do not occur until $t=10^9 T_{\rm K} 
\simeq 3 \times t_{\rm dep}$ whereas several giant impacts occur during the long term evolution
$(t \gtrsim 5 \times t_{\rm dep})$ for runs Ba1--5. A slight increase of $M_{\rm max}$ is anticipated
for run Da1 after $t > 10^9 T_{\rm K}$.

\subsection{Dependence on Disk Viscosity}
Many of the models in this paper, we assume that the disk viscosity is relatively small $(\alpha = 10^{-4})$.
However, as shown in OIK13, the evolution and final orbital configuration of
planets is quite different when $\alpha = 10^{-2}$ is considered.
All results for $\alpha = 10^{-2}$ are also summarized in Table~\ref{tbl:results}.

\begin{figure*}[htbp]
\epsscale{0.7}
\plottwo{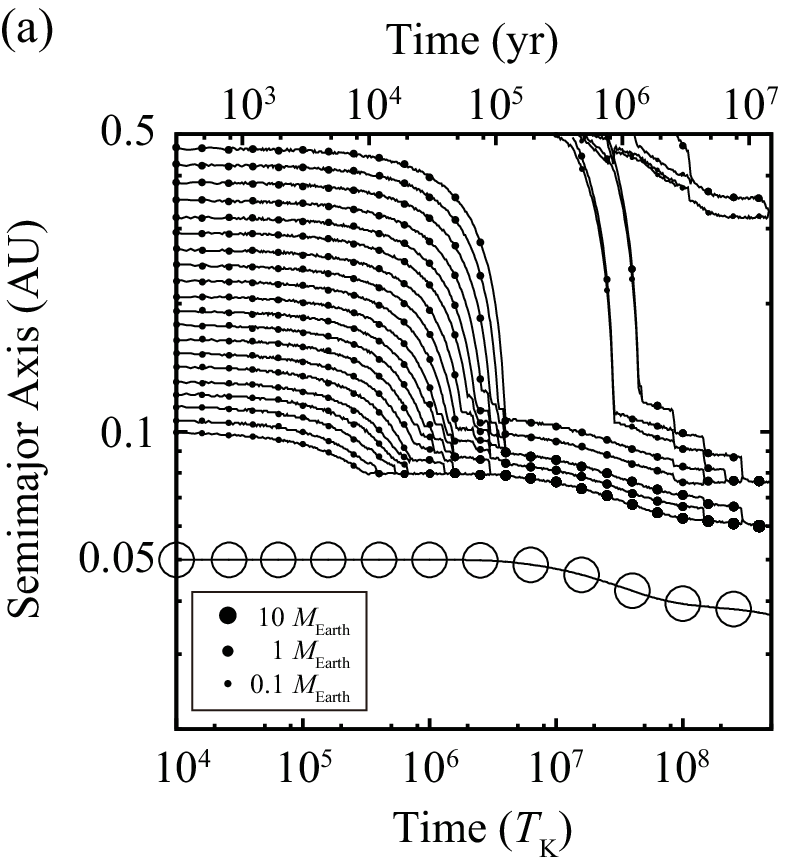}{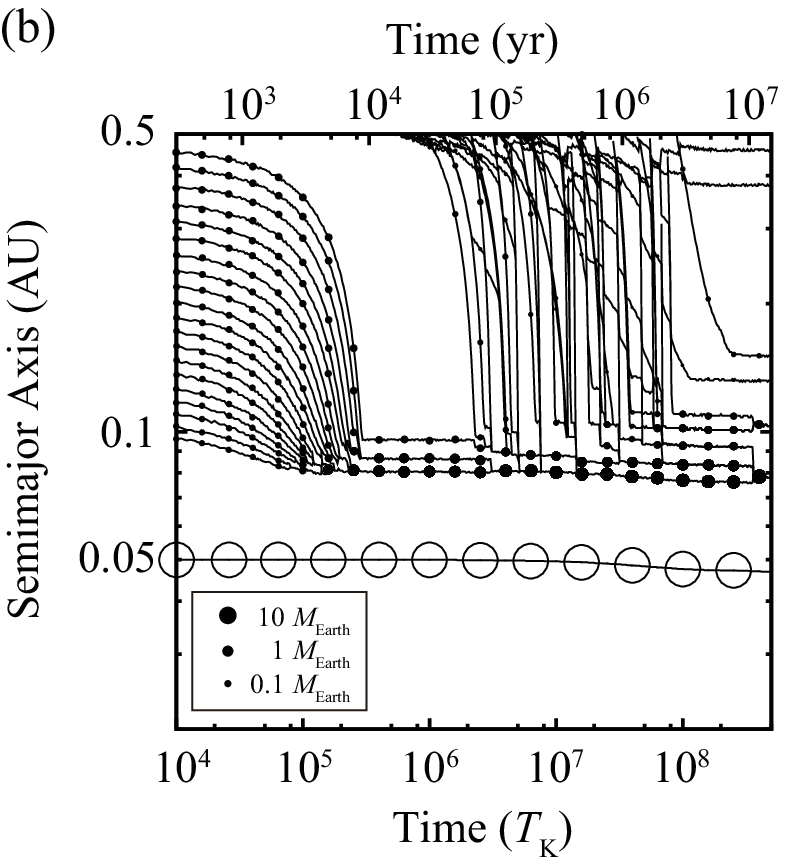}
\caption{Results of runs (a) Eb1 $(C_{\rm I}=0.1)$ and (b) Ec1 $(C_{\rm I}=10)$ simulations,
in which $\alpha = 10^{-2}$ is assumed as in OIK13.}
\label{fig:t_a_vis}
\end{figure*}

The typical orbital evolution is shown in Figure~2(a) of OIK13, which is the result of run Ea1.
The innermost planet in the resonant chain is captured in a 2:1 MMR
with the HJ because the location of the 2:1 MMR with the HJ is more distant from the
central star than that of the gap edge (Figure~\ref{fig:model}(b)).
Then, all planets in the resonant chain interact with the HJ, leading to
inward migration of the HJ. This induced migration of the HJ is called ``crowding--out'' by
smaller planets. The HJ is eventually lost in a collision with the central star. The efficiency of
crowding--out depends on the solid surface density in the disk; when $f_{\rm d}$ is small,
the HJ does not undergo induced migration.
It also depends on the efficiency of type I migration; crowding--out is not efficient enough
for small $C_{\rm I}$. In fact, in the results for $C_{\rm I}=0.1$ (runs Eb1--3), the HJ undergoes
little migration. Note that in the results for $C_{\rm I}=10$ (run Ec1--3), crowding--out
is also not observed (see Figure~\ref{fig:t_a_vis}(b)). 
This is because the innermost planet in the resonant chain, which
is located at the gap edge and does not yield a strong negative torque, has a large mass, 
and the total negative type I torque, which is suffered by other planets in the resonant 
chain, is too small to push the HJ inward efficiently. (In other words, the factor $\zeta$ in 
OIK13 is small.)
This mechanism of induced migration of HJs depends on the detailed structure of the gap
edge and should be investigated further.

\section{TORQUES ON PLANETS}
\label{sec:torques}
\subsection{Angular Momentum Transfer}

\begin{figure*}[htbp]
\epsscale{0.37}
\plotone{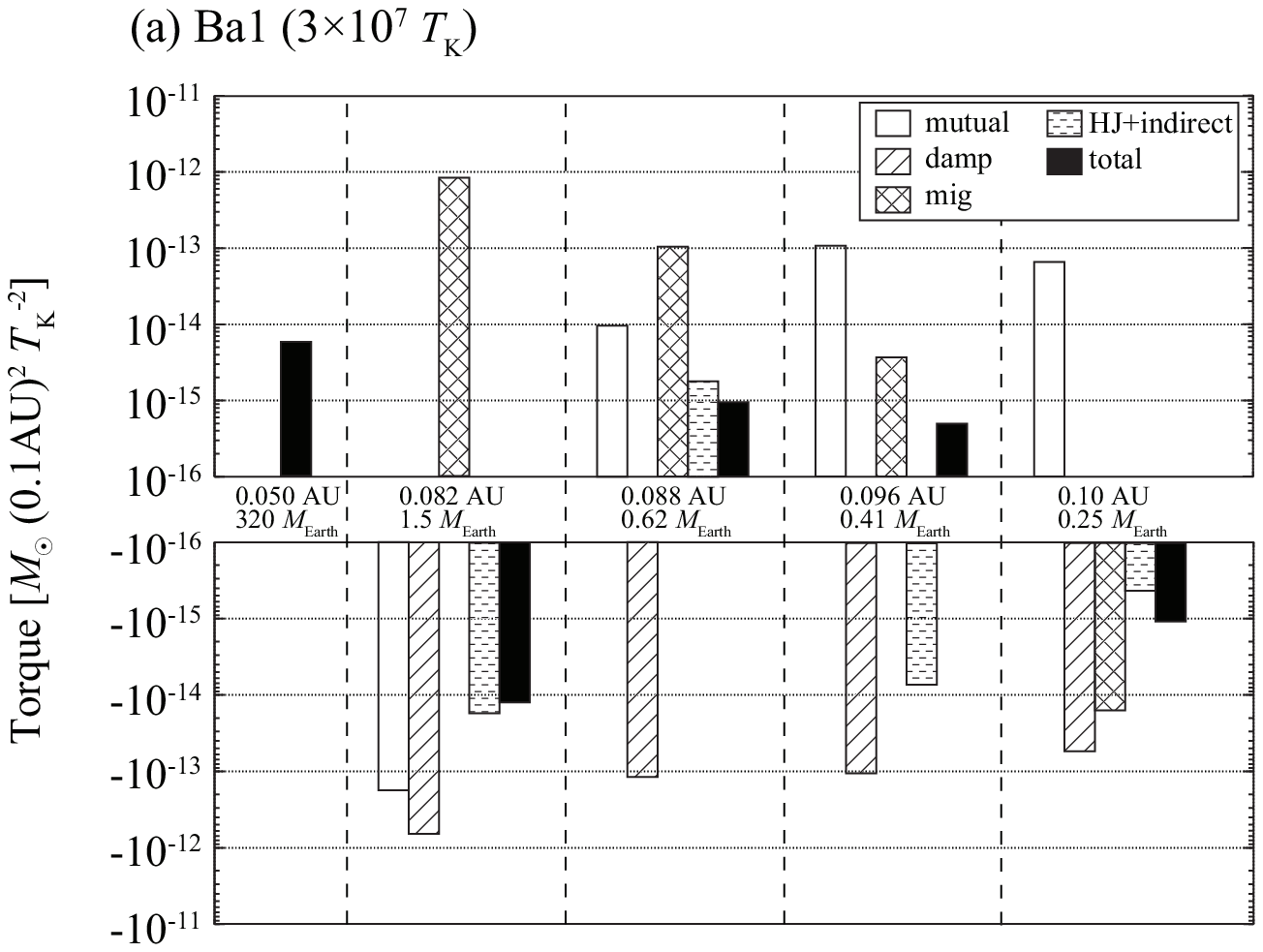}
\plotone{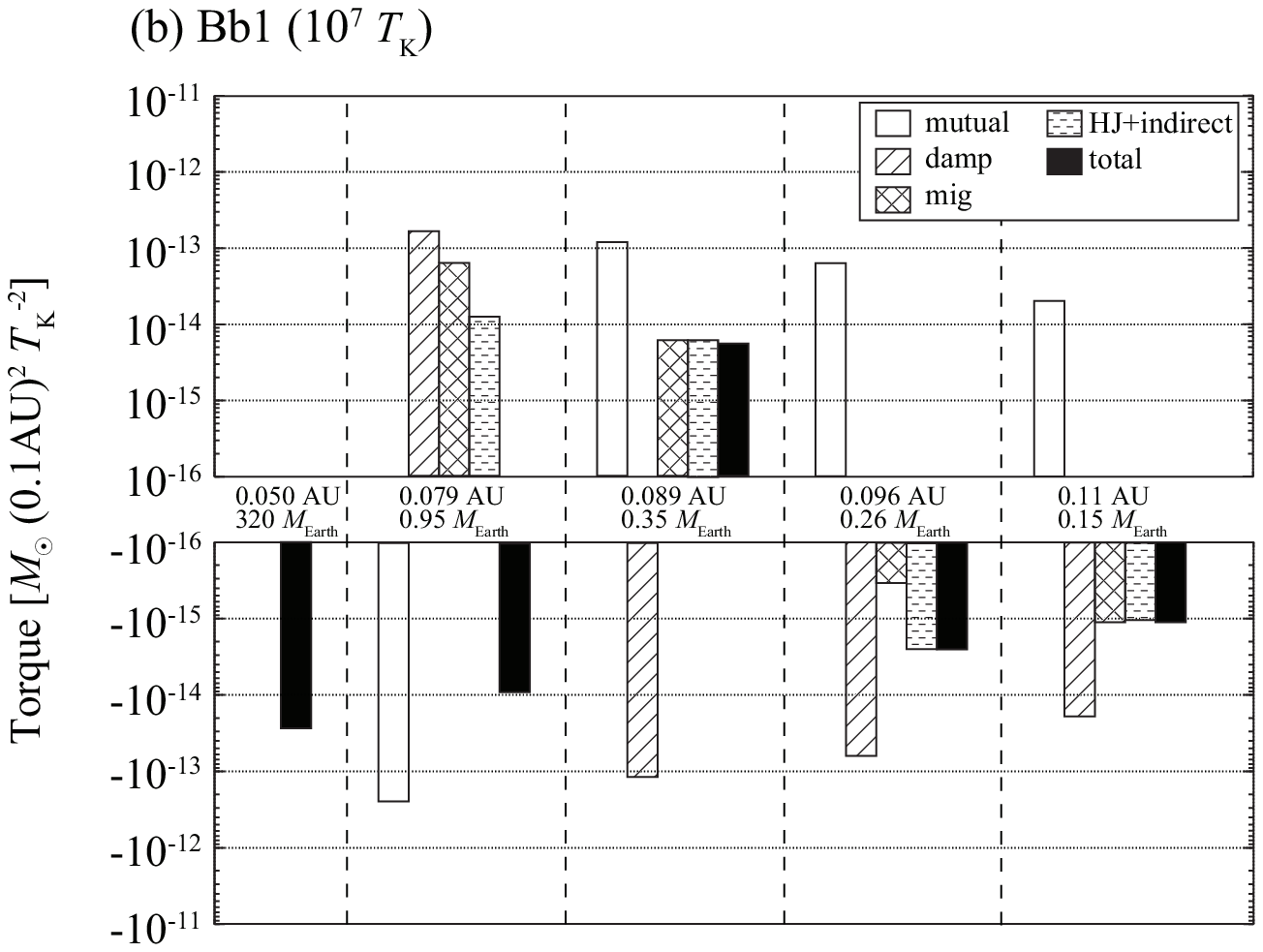}
\plotone{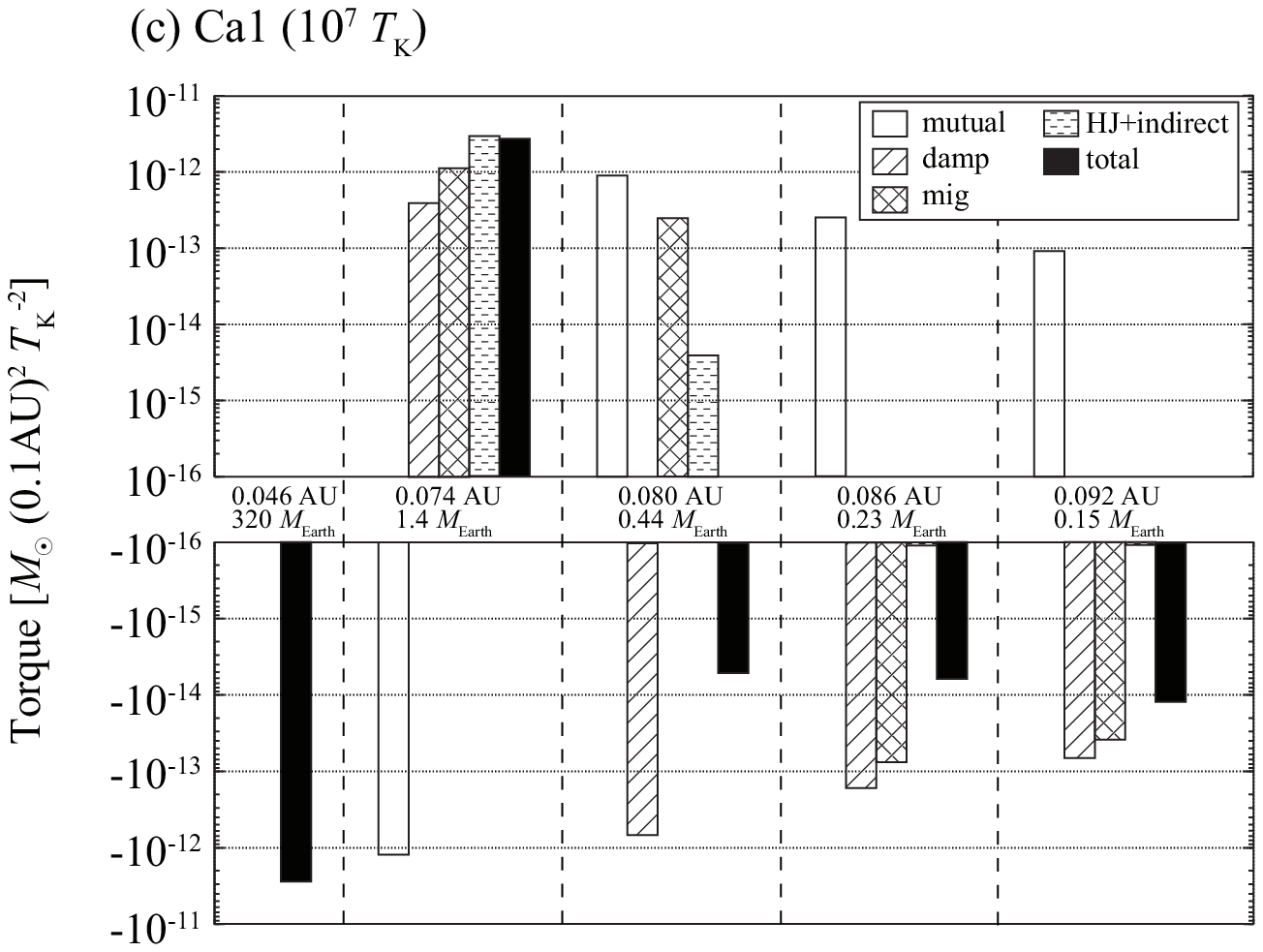}
\caption{Torques experienced by the HJ and innermost four planets for runs (a) Ba1 (fiducial
model), (b) Bb1 $(C_{\rm I}=0.1)$, and (c) Ca1 $(\alpha = 10^{-2})$. Black bars indicate
the total torque exerted on each planet, while the other bars represent each individual component
of the torque.}
\label{fig:a_T}
\end{figure*}

Here, we investigate angular momentum transfer in the system.
Figure~\ref{fig:a_T}(a) shows torques experienced by planets (the HJ and 
the innermost four planets)
for run Ba1 at $t = 3\times 10^7~T_{\rm K}$. Black bars indicate the total torque exerted on
the planets, while the other bars represent the mutual torque (sum of gravitational torques
from terrestrial planets), the damping torque ($e$--damping torque), the migration torque (type I migration
torque), and the HJ torque (sum of gravitational torque from the HJ and contributions from 
indirect terms).
The torques are normalized by $M_\odot r_0^2/t_0^2$ for $r_0 = 0.1 {\rm AU}$ and $t_0 = T_{\rm K}|_{a=0.1 {\rm AU}}$.
Each value is averaged over 1 Myr. The orbits of the planets in this system are 
almost steady at this moment, thus the torques on each body are equilibrated.

The sign of the type I migration torque depends on the slope of the surface density profile;
for run Ba1 at $3 \times 10^7 T_{\rm K}$,
the fourth innermost solid planet at $a=0.1~{\rm AU}$ loses angular momentum, while
the innermost three planets gain it from the type I migration torque. 
For the innermost planet, the magnitude of the positive type I torque is larger than that of the
negative $e$--damping torque, thus the excess angular momentum is transferred to the outer
planets through resonant interactions between planets.
As for the second innermost planet,
the positive type I torque almost balances the negative $e$-damping torque so that there is
only a small net mutual torque exerted on the planet. The transferred positive torque from the
innermost planet is predominantly
consumed by the third and fourth planets. We see that the angular momentum
transferred from/to the HJ is negligible.

\citet{ogihara_etal10} have found that when a planet with nonzero eccentricity resides 
near the inner edge of a disk, the planet gains angular momentum from the disk owing to
drag forces, $\textbf{\textit{F}}_{\rm damp}$, that are attributed to the velocity difference between 
the planet and the disk gas.
The positive torque exerted on the planet in the edge is called edge torque.
The magnitude of the edge torque depends on
the gas density difference between the pericenter and apocenter, in other words
the eccentricity of the planet and the sharpness of the edge, and the strength of the drag force.
In our model, the $e$--damping force acts as edge torque because it contains the term
corresponding to the difference between the orbital velocity of the planet and the gas velocity.

We see in Figure~\ref{fig:a_T}(a) that no edge torque operates on planets.
This is because the eccentricity of the innermost planet is relatively small ($\simeq 0.006$).
On the other hand, Figure~\ref{fig:a_T}(b) shows the torques on planets 
at $t = 10^7~T_{\rm K}$ for run Bb1, for which $C_{\rm I}=0.1$ is assumed.
We find that the innermost planet at $a = 0.079 {\rm AU}$ is subject to a positive 
$e$--damping torque, which is the edge torque. 
The eccentricity of the innermost planet is slightly excited to $\simeq 0.02$.

Figure~\ref{fig:a_T}(c) shows the torques on planets for run Ea1 at $t=10^7~T_{\rm K}$,
where the inward migration of the HJ, ``crowding--out'', is observed.  
The force acting on the HJ 
is primarily the gravitational force from the innermost planet, which is captured in a
2:1 resonance with the HJ. 
This indicates a transfer of angular 
momentum from the HJ to the planets in the resonant chain.

\subsection{Torque due to the Velocity Difference between a Planet and Gas}
\label{sec:velocity_difference}
We find that the eccentricity damping force gives a contribution to the 
change of the semimajor axis. 
From $\textbf{\textit{F}}_{\rm damp}$, we can calculate $da/dt$ averaged over
an orbit as 
\begin{eqnarray}
a^{-1} \frac{da}{dt} = \frac{e^{2}}{0.78 t_{e}} (A^{s}_{r} + A^{c}_{\theta}) + \mathcal{O}(e^3),
\end{eqnarray}
where $A^{s}_{r} + A^{c}_{\theta} = -0.69$. If the eccentricity is small enough,
this migration rate is negligible. However, eccentricities are pumped up by mutual 
interactions between planets, resulting in planetary migration caused by 
$\textbf{\textit{F}}_{\rm damp}$.

In the above discussion, we assume that planets follow a Keplerian orbit, however
they can be changed by the gravitational influence of the HJ. 
In many runs of the simulation, we observe that the orbital velocity near the
apocenter is not so slow that the positive tailwind torque near the apocenter cannot compensate
for the negative headwind torque near the pericenter, resulting in a net negative
torque (inward migration). This arises because the orbital velocity of the planet
is altered by the gravity of the HJ.

This phenomenon has several characteristics. When the perturber resides  
inside the orbit of the planet, the velocity of the planet is increased leading to 
a loss of angular momentum via gas drag. On the other hand, if the perturber
lies outside the orbit of the planet, the orbital velocity of the planet is decreased and
the planet gains angular momentum, resulting in a net outward migration.
The magnitude of the change in the orbital velocity depends both on the mass of the
perturber and its distance from the planet.
A more detailed study of this phenomenon should be done in future work.

\section{PROPERTIES AND PARAMETER DEPENDENCE OF PLANETARY SYSTEMS}
\label{sec:discussion}

We discuss several of the final properties of planetary systems and their dependence on
parameters.

\subsection{Mass of Planets}
\label{sec:mass}

The typical masses of the largest planets and the total mass in planets are $1-2 M_\oplus$
and $2-4 M_\oplus$, respectively, and depend little on $C_{\rm I}$ but strongly on $f_{\rm d}$.
The mass ratio between the largest planet and the total mass is typically $\simeq 0.4-0.8$,
whereas the value decreases down to $\simeq 0.1$ for $f_{\rm g}=0.01$.

In OIK13, we derived the total mass in the resonant chain, $M_{\rm chain}$,
(see section~4 in OIK13) where $M_{\rm chain}$ is the total mass of planets
that migrate to the vicinity of the gap edge before gas depletion. Thus, 
$M_{\rm chain}$ is comparable to $M_{\rm tot}$ and also correlates with $M_{\rm max}$.
According to OIK13, $M_{\rm chain}$ can be expressed in two different ways
depending on the values of parameters. As stated in Section~\ref{sec:results1}
and OIK13, the mass of migrating planets is given by the critical mass 
for migration or the isolation mass, whichever is smaller. The condition where the mass of
all the migrating planets is expressed by the isolation mass is
\begin{eqnarray}
&&2 \left(\frac{t_{\rm dep}}{10^6 {\rm yr}}\right)^{4/3}
f_{\rm d}^{27/11} C_{\rm I}^{59/33} f_{\rm g}^{53/33}
\nonumber \\&&\times 
\left(\frac{M_*}{M_\oplus}\right)^{-5/33}
\left(\frac{L_*}{L_\odot}\right)^{-1/3} \la 1.
\end{eqnarray}
When $f_{\rm d}$, $f_{\rm g}$, and $C_{\rm I}$ are small, this inequality is satisfied.
In that case, the total mass in the resonant chain is
\begin{eqnarray}
\label{eq:m_chain_iso}
M_{\rm chain, iso} &\simeq& 0.1 \left(\frac{t_{\rm dep}}{10^6 {\rm yr}}\right)^{2/3}
\left(\frac{f_{\rm d}}{0.1}\right)^{3}
C_{\rm I}^{2/3} f_{\rm g}^{2/3}
\nonumber \\&&\times
\left(\frac{L_*}{L_\odot}\right)^{-1/6} 
M_\oplus .
\end{eqnarray}
In the other case, where the mass of the migrating bodies is expressed by the critical
mass, the total mass is given by
\begin{eqnarray}
\label{eq:m_chain_crit}
M_{\rm chain, crit} &\simeq& 5.6 \left(\frac{t_{\rm dep}}{10^6 {\rm yr}}\right)^{5/24}
f_{\rm d}^{37/32} C_{\rm I}^{5/96} f_{\rm g}^{11/96}
\nonumber \\&&\times
\left(\frac{M_*}{L_\odot}\right)^{5/96}
\left(\frac{L_*}{L_\odot}\right)^{-5/96} 
M_\oplus .
\end{eqnarray}
The final masses can be roughly estimated using Equations~(\ref{eq:m_chain_iso}) and 
(\ref{eq:m_chain_crit}). 
Note that in most simulations in this study, $M_{\rm chain}$ is expressed using
Equation~(\ref{eq:m_chain_crit}). We find from this equation that the mass is weakly
dependent on $C_{\rm I}$ and is roughly linearly dependent on $f_{\rm d}$.
$M_{\rm chain}$ can be regarded as $M_{\rm tot}$; for example, the total mass 
in planets is $\simeq 3.4 M_\oplus$ for runs Ba1--5 and $\simeq 25 M_\oplus$ for runs Be1--5,
consistent with the estimate of Equation~(\ref{eq:m_chain_crit}).

The typical mass ratio  of $M_{\rm max}/M_{\rm tot} \simeq 0.5$ means that one large
planet dominates the mass in planets. Migrating protoplanets formed via oligarchic growth
have comparable masses, but are blocked at the traps (e.g., gap edge) and collide with
one another, resulting in mass accumulation  near the traps. Thus, one can argue that
the gap and the 2:1 resonance act as traps that concentrate solid materials in one planet.
During this process, orbital rearrangements of the resonant chain 
caused by planets that migrate from the
outer region induce collisions between planets, which increases the fraction of $M_{\rm max}$.
In other words, when the migration is less effective and orbital rearrangement hardly
occurs, $M_{\rm max}/M_{\rm tot}$ can also be kept small.
For example, in the results for $f_{\rm g}=0.01$, the $a$--damping rate (type I migration and
$e$--damping) is small, which results in smaller values of $M_{\rm max}/M_{\rm tot} \simeq 0.1$.

\subsection{Resonant Configurations}
\label{sec:resonance}
At the gap edge, planets no longer migrate and capture other planets migrating
from the outer disks into resonances.
The typical values of resonant commensurabilities are between 4:3 and 9:8, and 
have little dependence on the parameters. The condition for capture into first-order 
MMRs among two bodies is derived by \citet{ogihara_kobayashi13}. 
The critical migration
timescale for capture into a $p+1:p$ resonance, where $p$ is an integer, is
\begin{equation} 
t_{a,{\rm crit},p} = C_p \left(\frac{M}{M_{\oplus}}\right)^{4/3} 
\left(\frac{M_{*}}{M_{\odot}}\right)^{-4/3} T_{\rm K},
\label{eq:ta_crit}
\end{equation}
where the values of the numerical coefficients are $C_1 = 1\times 10^7$,
$C_2 = 5\times 10^5$, $C_3 = 2\times 10^5$, and $C_4 = 1\times 10^5$ between
small and large planets.
Note that for resonance capture between planets with comparable masses, the coefficients
are changed (see Table~2 of \citealt{ogihara_kobayashi13}).
If the relative migration timescale between two bodies is between the critical migration
timescales $t_{a,{\rm crit},p-1}$ and $t_{a,{\rm crit},p}$, capture into a
$p+1:p$ resonance is assured unless the eccentricity at the resonant encounter 
is not sufficiently high $(e \lesssim 0.1)$. 
Although planets in this study are captured in multiple MMRs, we can roughly discuss the 
origins of resonant configurations with the use of $t_{a,{\rm crit}}$ for two--body resonances.

For runs Ba1--5 at $t = 10^7 T_{\rm K}$, the mass of the body migrating from the outer region 
is $\simeq 0.1 M_\oplus$
and its migration timescale is $\simeq 5 \times 10^5 T_{\rm K}$. The maximum planetary mass
in the resonant chain is $\simeq 1 M_\oplus$, thus the mass ratio between the migrating body
and the largest planet is $\simeq 0.1$. 
From Equation~(\ref{eq:ta_crit}), the planet is expected to be captured in a 3:2 or 4:3 
resonance. Indeed, several planets are captured in 4:3 resonances, while some other
pairs are captured in high $p$ resonances due to gravitational pushing by other planets.

We can discuss resonant capture by comparing $t_a$ and $t_{a,{\rm crit},p}$.
When $C_{\rm I}$ is altered between 0.1 and 10, both the mass of the largest planet and 
the migration timescale are not changed significantly, as described in Section~\ref{sec:mass}, 
thus $t_{a}/t_{a,{\rm crit},p} = 5 \times 10^5/C_{p}$,
which is almost independent of $C_{\rm I}$.
Therefore, the resonant configuration is insensitive to $C_{\rm I}$.

Similarly, we can discuss the dependence on $f_{\rm d}$.
For the case of $f_{\rm d}=10$, the migration timescale of migrating bodies is
$t_a \simeq 5 \times 10^4 T_{\rm K}$, and the mass of the largest planet is
 $M \simeq 10 M_\oplus$, therefore $t_{a}/t_{a,{\rm crit}} = 10^{6}/C_{p}$.
Because this value is slightly larger than that for the fiducial model,
the resonant configuration is in a slightly separated resonance, such as 3:2
(see Table~\ref{tbl:results} for runs Be1--5).
For the case of $f_{\rm d} = 0.1$, both $t_{a}$ and $t_{a,{\rm crit}}$ are increased by a
factor of $\simeq 10$, and therefore the resonant commensurabilities are almost
the same as those for the case of $f_{\rm d} = 1$.

In short, resonant configurations can be discussed with use of $t_{a,{\rm crit}}$. 
The migration timescale is basically shorter than $t_{a,{\rm crit},1}$ (the critical
migration timescale for a 2:1 resonance), 
which means that closely spaced resonances (e.g., 5:4) are
favored. In addition, we find that the formation of the chain of resonant planets
is a robust process. 
In our series of simulations with a range of input parameters, in which the growth
and migration are simultaneously followed, the migration timescale of planets
which migrate from the outer region is always longer than $10^4 T_{\rm K}$,
therefore planets can be captured in some closely spaced resonances
as shown by \citet{ogihara_kobayashi13}.
Note that as observed in the results of long--term calculations (e.g., run Ba1), resonant
configurations can be lost during the long--term evolution; although the formation of
resonances is a robust process, they do not necessarily remain until the end.

In the above discussion, the condition for capture into two--body MMRs
is used, however, the use of this formula is not strictly appropriate when one
considers capture into multiple MMRs. 
This is an issue that we will address in a future publication.

\subsection{Other Properties}
As for the orbital separations between
planets, the smallest separation for each run is typically below 10 Hill radii.
According to an investigation of orbital stability of non--resonant multiple planet systems
\citep{chambers_etal96}, such systems with relatively small separations should 
become unstable within a timescale of $\simeq 10^8 T_{\rm K}$. However,
we find that most systems are stable after $5 \times 10^8 T_{\rm K}$:
this is because the orbital crossing time of planets in a resonant chain 
can become significantly longer than that of non--resonant planets
\citep{matsumoto_etal12}. 

\section{LACK OF COMPANION PLANETS IN HOT JUPITER SYSTEMS}
\label{sec:observation}
Observations have suggested that there is a lack of companion planets 
in HJ systems (e.g., \citealt{steffen_etal12}).
In OIK13, we performed \textit{N}--body simulations of planet formation
assuming $\alpha = 10^{-2}$ and found that 
several terrestrial planets efficiently form outside the orbit of an HJ. They gravitationally
interact with the HJ through resonances, which leads to the inward migration of the HJ
(crowding--out). We proposed that our model naturally explains the lack of
additional observed planets in orbits close to HJs regardless of the disk mass.
When planet formation occurs in a less massive disk, crowding--out is not effective and
the HJ and terrestrial planets can coexist. However, the companion planets tend to remain 
small and below the detection limit of current observations.

For $\alpha = 10^{-4}$, we find that several terrestrial 
planets are robustly formed outside the orbit of an HJ and captured in a resonant chain, and
they hardly interact with the HJ. As a result, both the terrestrial planets and HJ coexist 
at the end. In addition, $M_{\rm max}/M_{\rm tot}$ is high $(\simeq 0.5)$, thus a few large
terrestrial planets tend to form. Therefore, it seems likely that terrestrial planets could be
observed in orbits close to HJs by current surveys, and that 
future observations may detect these systems. However, because observational data
obtained so far clearly indicate the lack of companion planets in HJ systems, 
it suggests that turbulent viscosity is larger than the value adopted $(\alpha = 10^{-4})$
because the region in the vicinity of the central star is considered to be MRI active.
It is also possible that terrestrial planets pass through the gap edge.

Although we found that planetary systems that consist of only terrestrial planets
can be reproduced by crowding--out of HJs, it is worth mentioning that 
the crowding--out has not necessarily been experienced by such systems. 
According to \textit{N}--body 
simulations in which formation of the terrestrial planets is investigated without considering
HJs, multiple close--in terrestrial planets form captured in MMRs (e.g., \citealt{terquem_papaloizou07}).
It has also been demonstrated that multiple terrestrial planets that are not locked in MMRs 
form if the type I migration speed is reduced by a factor of 100 from that predicted by 
the linear theory \citep{ogihara_ida09}.
To examine various models,
it is important to compare predictions from formation models with observational data statistically; 
however, we leave this for subsequent work.

\section{CONCLUSIONS}
\label{sec:conclusions}
We have investigated planetary accretion in the presence of an HJ with a range of
various model parameters. Through detailed \textit{N}--body calculations, we confirm the
following planet formation process.
\begin{enumerate}
\item Embryo formation stage: the first stage after formation of planetesimals is the oligarchic
growth stage. Planetesimals grow to protoplanets while keeping their mutual orbital separations
$\simeq 8 r_{\rm H}$. This stage terminates when the embryo mass reached the isolation mass
or if the embryos begin to migrate faster. The typical duration of this stage is
$10^5~T_{\rm K} \simeq 3 \times 10^3~{\rm yr}$.
\item Migration stage: the protoplanets undergo inward migration with the migration
timescale $t_a$, which is determined by the embryo mass and
several parameters (e.g., $f_{\rm g}$ and $C_{\rm I}$). The innermost body ceases its migration
when it is trapped by the gap edge or captured in a 2:1 MMR with the 
HJ, whichever happens first. During this stage, embryos that formed in the outer region
sequentially migrate inward before the depletion of the gas disk. They are captured in MMRs
with inner planets that reside in a chain of resonant planets or undergo
close encounters with the inner bodies leading to a rearrangement of orbital configurations.
The innermost planet tends to be the largest body, which consists of about 50 percent of the total
mass of the close--in planets. When the resonant
chain is captured in a 2:1 resonance with the HJ, the HJ can be pushed 
inward to the vicinity of the central star, the migration speed of which depends on the
conditions. This stage continues until the gas has almost decayed $(t \la 3 t_{\rm dep})$.
\item Final stage: after the gas depletion, final orbital configurations are established.
In some cases, as the effect of eccentricity--damping weakens,
planets exhibit local orbital crossings resulting in giant impacts. Even after local
orbital instability, several commensurabilities tend to remain, although in some cases all
resonant relations are lost via collisions.
\end{enumerate}

We have also determined the dependence of our results on model parameters.
Several properties of the final states can be summarized as follows.
\begin{description}
\item[Mass:] The typical mass of the largest planet is $\simeq 1-2~M_\oplus$. The mass ratio
of the largest planet and the total mass in planets is large $(\simeq 0.4-0.8)$.
Owing to the stalling of migration, solid bodies are accreted by one or a few planets,
and eventually a large difference in mass between the largest planet and small planets is created. 
The mass also hardly depends on $C_{\rm I}$.  An increase/decrease in
$f_{\rm d}$ results in an increase/decrease of the embryo growth rate and mass.
\item[MMR:] We find that the formation of chains of resonant planets is robust if both
migration and trapping by the edge or the 2:1 resonance with the HJ are effective.
The typical commensurability in the resonant chain is between the 4:3 resonance
and the 9:8 resonance, which also weakly depends on the input parameters 
(e.g., $C_{\rm I}$ and $f_{\rm d}$). 
We also observe a tendency that orbital crossings after gas dissipation occur more frequently
in the case of large $f_{\rm d}$. Almost no planets formed are captured in 1:1 coorbital resonances.
\end{description}

Through a series of simulations, our understanding of the reason for the lack of companion
planets in HJ systems is improved. If $\alpha = 10^{-4}$ is assumed, it is difficult to
explain the observations. Therefore, this may suggest that the disk viscosity 
near the HJ is high $(\alpha \simeq 10^{-2})$ and the crowding--out of the HJ by 
terrestrial planets is effective.

This study provides several suggestions for future study. One is the fact that the results for 
the final orbital configuration are almost identical for  small--$N$ and  large--$N$ simulations. 
Thus, for the purpose of examining 
the configuration of final states, \textit{N}--body simulations can be started with relatively
large protoplanets. 
In addition, we show capability of numerical simulations
that combine an \textit{N}--body code with a semi--analytic population synthesis model.
We also find the orbital velocity of planets that are located near the HJ 
to be altered, resulting in the transport of angular momentum through gas drag.
This should be investigated in detail using high--resolution dynamical simulations in which 
the gas motion is also calculated by a magneto--hydrodynamical code.
This study can be applied to the formation of planets outside warm/cool Jupiters, which
may solve several issues regarding the formation of gaseous/icy planets in the Solar system.
As stated in Sections~\ref{sec:intro} and \ref{sec:model}, it is of particular importance to resolve
the gas flow around HJs and to examine the structure and evolution of the inner disk, which
should be investigated in future studies.

\subsection*{ACKNOWLEDGMENT}
We thank the anonymous referee for useful comments.
We also thank Shoichi Oshino, Eiichiro Kokubo, and Yasunori Hori for helpful discussions, 
and thank Jennifer M. Stone for variable comments.
Numerical computations were in part conducted on the GRAPE system and
general-purpose PC farm at the Center for Computational Astrophysics, 
CfCA, of the National Astronomical Observatory of Japan.
This work is supported by a Grant-in-Aid for JSPS Fellows (23004841).

{}

\LongTables
\begin{deluxetable*}{lcccclc}
\tabletypesize{\tiny}
\tablecolumns{8}
\tablewidth{0pc}
\tablecaption{Simulation results}
\startdata
\hline \hline
Run		& \parbox[m]{2em}{\strut $M_{\rm max}$ $(M_\oplus)$\strut} &	  \parbox[m]{2em}{\strut $M_{\rm tot}$ $(M_\oplus)$\strut} 	&  $M_{\rm max}/M_{\rm tot}$	& $N$	& Commensurability	& $t (T_{\rm K})$\\
\hline
Aa1	& 1.56 & 3.20	& 0.49	& 6	& 10:9(1--2), 9:8(2--3), 8:7(3--4)& $5\times 10^8$\\
		& 1.56 & 3.20	& 0.49	& 6	& 10:9(1--2), 9:8(2--3), 8:7(3--4)& $1\times 10^9$\\
Aa2		& 2.38 & 3.30	& 0.72	& 5	& 3:2(2--3), 9:8(4--5)& $5\times 10^8 $\\
		& 2.38 & 3.30	& 0.72	& 5	& 3:2(2--3), 9:8(4--5)& $1\times 10^9 $\\
Aa3		& 3.04 & 3.32	& 0.92	& 3	& none	& $5\times 10^8 $\\
		& 3.04 & 3.32	& 0.92	& 3	& none	& $1\times 10^9 $\\
Ba1		& 1.46 & 3.34	& 0.44	& 6	& 7:6(1--2), 9:8(3--2) & $5\times 10^8 $\\
		& 2.99 & 3.34	& 0.90	& 4	& none & $1\times 10^9 $\\
Ba2		& 1.38 & 3.37	& 0.41	& 5	& 7:6(1--2), 7:6 (2--3), 2:1(3--4) & $5\times 10^8 $\\
		& 3.24 & 3.37	& 0.96	& 3	& none & $1\times 10^9 $\\
Ba3		& 2.52 & 3.28	& 0.77	& 3	& 5:4(1--2) & $5\times 10^8 $\\
		& 2.52 & 3.28	& 0.77	& 3	& 5:4(1--2) & $1\times 10^9 $\\
Ba4		& 2.66 & 3.50	& 0.76	& 4	& 4:3(1--2) & $5\times 10^8 $\\
		& 2.66 & 3.50	& 0.76	& 4	& 4:3(1--2) & $1\times 10^9 $\\
Ba5		& 1.51 & 3.39	& 0.45	& 8	& 7:6(1--2), 4:3(2--3), 9:8(3--4), 12:11(4--5) & $5\times 10^8 $\\
		& 2.63 & 3.39	& 0.78	& 5	& 4:3(2--3) & $1\times 10^9 $\\
Bb1		& 1.57 & 2.40	& 0.65	& 6	& 4:3(1--2) & $5\times 10^8 $\\
Bb2		& 1.43 & 2.34	& 0.61	& 8	& 7:6(3--4), 6:5(4--5), 9:8(6--7) & $5\times 10^8 $\\
Bb3		& 0.96 & 2.40	& 0.40	& 8	& 2:1(0--1), 8:7(1--2), 7:6(2--3), 9:8(3--4), 11:10(5--6) & $5\times 10^8 $\\
Bb4		& 1.29 & 2.55	& 0.51	& 9	& 5:4(3--4), 7:6(4--5), 9:8(8--9) & $5\times 10^8 $\\
Bb5		& 1.57 & 2.33	& 0.67	& 7	& 6:5(2--3), 5:4(3--4), 11:10(4--5), 11:10(6--7) & $5\times 10^8 $\\
Bc1		& 1.48 & 3.70	& 0.40	& 7	& 8:7(1--2), 7:6(2--3), 8:7(3--4), 9:8(4--5), 4:3(5--6) & $5\times 10^8 $\\
Bc2		& 2.29 & 3.53	& 0.65	& 8	& 9:8(1--2), 9:8(2--3), 6:5(3--4) & $5\times 10^8 $\\
Bc3		& 1.65 & 3.63	& 0.45	& 10	& 9:8(1--2), 7:6(2--3), 8:7(3--4), 6:5(4--5), 5:4(5--6), 3:2(7--8) & $5\times 10^8 $\\
Bc4		& 1.63 & 3.66	& 0.45	& 8	& 7:6(1--2), 5:4(2--3), 7:6(4--5), 12:11(5--6), 8:7(6--7) & $5\times 10^8 $\\
Bc5		& 2.31 & 3.60	& 0.64	& 7	& 8:7(1--2), 6:5(2--3), 7:6(4--5), 9:8(5--6) & $5\times 10^8 $\\
Bd1		& 0.122 & 0.234	& 0.52	& 14	& 8:7(1--2), 7:6(2--3), 8:7(4--5), 6:5(5--6), 6:5(7--8), 9:8(9--10) & $5\times 10^8 $\\
Bd2		& 0.0988 & 0.229	& 0.43	& 17	& 9:8(1--2), 10:9(3--4), 8:7(4--5), 7:6(7--8) & $5\times 10^8$\\
Bd3		& 0.0685 & 0.235	& 0.29	& 18	& 14:13(1--2), 7:6(2--3), 11:10(3--4), 10:9(4--5), 6:5(5--6) & $5\times 10^8$\\
Bd4		& 0.115 & 0.233	& 0.49	& 16	& 8:7(1--2), 9:8(2--3), 9:8(3--4), 9:8(4--5), 8:7(6--7), 6:5(8--9) & $5\times 10^8$\\
Bd5		& 0.115 & 0.229	& 0.50	& 15	& 6:5(2--3), 9:8(3--4), 7:6(7--8), 2:1(8--9) & $5\times 10^8$\\
Be1		& 11.5 & 25.6	& 0.45	& 7	& 3:2(1--2), 5:4(3--4), 4:3(4--5), 6:5(5--6) & $5\times 10^8 $\\
Be2		& 19.7 & 25.0	& 0.79	& 2	& none & $5\times 10^8 $\\
Be3		& 13.0 & 24.9	& 0.52	& 4	& 2:1(0--1), 3:2(1--2), 4:3(3--4) & $5\times 10^8 $\\
Be4		& 15.2 & 25.5	& 0.60	& 6	& 2:1(0--1) & $5\times 10^8 $\\
Be5		& 15.2 & 25.4	& 0.60	& 3	& 3:2(1--2) & $5\times 10^8 $\\
Ca1		& 0.223 & 1.88	& 0.12	& 14	& 3:2(1--2), 6:5(3--4), 4:3(5--6), 8:7(7--8), 7:6(9--10) & $5\times 10^8 $\\
Ca2		& 0.196 & 1.88	& 0.10	& 14	& 4:3(1--2), 5:4(2--3), 5:4(3--4), 7:6(4--5), 8:7(5--6), 8:7(8--9) & $5\times 10^8 $\\
Ca3		& 0.261 & 1.88	& 0.14	& 11& 7:6(3--4), 5:4(7--8), 7:6(8--9), 7:6(9--10) & $5\times 10^8 $\\
Cb1		& 1.13 & 2.25	& 0.50	& 8	& 4:3(1--2), 9:8(3--4), 10:9(5--6) & $5\times 10^8 $\\
Cb2		& 1.06 & 2.21	& 0.48	& 10	& 9:8(3--4), 12:11(9--10) & $5\times 10^8 $\\
Cb3		& 1.29 & 2.26	& 0.57	& 8	& 11:10(5--6), 7:6(6--7) & $5\times 10^8 $\\
Cc1		& 3.18 & 4.58	& 0.69	& 5	& 9:8(3--4) & $5\times 10^8 $\\
Cc2		& 2.97 & 4.47	& 0.66	& 4	& 3:2(2--3) & $5\times 10^8 $\\
Cc3		& 3.62 & 4.60	& 0.79	& 4	& none & $5\times 10^8 $\\
Da1		& 1.88 & 4.56	& 0.39	& 6	& 8:7(1--2), 7:6(2--3), 9:8(3--4), 6:5(4--5), 3:2(5--6)	& $1\times 10^9$\\
Da2		& 1.80 & 4.85	& 0.37	& 5	& 6:5(1--2), 5:4(2--3), 5:4(3--4) & $1\times 10^9$\\
Da3		& 3.01 & 4.76	& 0.63	& 3	& none	& $1\times 10^9$\\
Db1		& 1.57	& 2.83	& 0.55	& 3	& none	& $1\times 10^9$\\
Db2		& 1.74	& 2.86	& 0.61	& 5	& 6:5(3--4), 4:3(4--5)	& $1\times 10^9$\\
Db3		& 1.72	& 2.77	& 0.62	& 5	& 6:5(1--2), 11:10(2--3), 14:13(4--5)	& $1\times 10^9$\\
Dc1		& 2.96	& 4.24	& 0.70	& 8	& 9:8(1--2), 7:6(2--3), 7:6(4--5) 	& $1\times 10^9$\\
Dc2		& 2.47	& 4.38	& 0.56	& 8	& 8:7(1--2), 7:6(2--3), 6:5(3--4), 5:4(4--5), 8:7(5--6)	& $1\times 10^9$\\
Dc3		& 3.24	& 4.42	& 0.73	& 6	& 6:5(1--2), 4:3(2--3), 3:2(3--4), 10:9(4--5)	& $1\times 10^9$\\
Dd1		& 0.178	& 0.283	& 0.63 	& 16	& 6:5(1--2), 13:12(8--9)	& $1\times 10^9$\\
Dd2		& 0.186	& 0.279	& 0.67	& 13	& 13:12(5--6), 8:7(9--10)	& $1\times 10^9$\\
Dd3		& 0.0892	& 0.283	& 0.32	& 18	& 10:9(3--4), 9:8(5--6),	13:12(8--9), 11:10(9--10)& $1\times 10^9$\\
De1		& 11.5 & 25.6	& 0.45	& 7	& 5:4(1--2), 6:5(2--3), 5:4(3--4), 5:4(4--5), 3:2(5--6), 8:7(6--7)& $1\times 10^9$\\
De2		& 15.2 & 25.2	& 0.60	& 3	& 3:2(1--2)	& $1\times 10^9$\\
De3		& 16.6 & 25.4	& 0.65	& 4	& 3:2(1--2)	&$1\times 10^9$\\
Ea1		& 2.29 & 3.36	& 0.68	& 2	& none & $5\times 10^8 $\\
Ea2		& 3.36 & 3.36	& 1.00	& 1	&  none & $5\times 10^8 $\\
Ea3		& 2.51 & 3.34	& 0.75	& 4	& 3:2(1--2)	& $5\times 10^8 $\\
Eb1		& 1.57 & 2.34	& 0.67	& 3	& none  & $5\times 10^8 $\\
Eb2		& 1.57 & 2.40	& 0.65	& 6	& 4:3(2--3) & $5\times 10^8 $\\
Eb3		& 1.19 & 2.40	& 0.50	& 4	& 5:4(0--1) & $5\times 10^8 $\\
Ec1		& 3.23 & 3.53	& 0.92	& 6	& 3:2(1--2), 6:5(3--4) & $5\times 10^8 $\\
Ec2		& 2.32 & 3.70	& 0.63	& 6	& 6:5(1--2), 5:4(2--3), 6:5(3--4), 3:2(4--5) & $5\times 10^8 $\\
Ec3		& 2.38 & 3.63	& 0.66	& 6	& 5:4(1--2), 8:7(5--6) & $5\times 10^8 $\\
Ed1		& 0.110 & 0.23	& 0.48	& 13	& 2:1(0--1), 8:7(1--2), 10:9(2--3), 7:6(3--4) & $5\times 10^8 $\\
Ed2		& 0.0988 & 0.23	& 0.43	& 14	& 2:1(0--1), 8:7(1--2), 8:7(2--3), 8:7(3--4), 5:4(4--5), 6:5(5--6), 7:6(6--7) & $5\times 10^8 $\\
Ed3		& 0.0788 & 0.23	& 0.34	& 18	& 2:1(0--1), 9:8(1--2), 9:8(2--3), 9:8(3--4), 10:9(4--5), 6:5(6--7) & $5\times 10^8 $\\
Ee1		& 20.6 & 23.6	& 0.87	& 3	& 3:2(1--2) & $5\times 10^8$\\
Ee2		& 17.4 & 23.6	& 0.74	& 2	& none	& $5\times 10^8 $\\
Ee3		& 14.2 & 23.6	& 0.60	& 2	& 3:2(1--2) & $5\times 10^8$
\enddata
\tablecomments{The variable $M_{\rm max}$ is the mass of the largest planet, $M_{\rm tot}$ is the
total mass in planets, and $N$ is the number of planets. The sixth column shows resonant states. For example, in run Aa1, the first innermost pair (the first and second innermost terrestrial planets), the second
innermost pair, and the third innermost pair are in 10:9, 9:8, and 8:7 MMRs, respectively. Indicated
mean motion commensurabilities apply to within 1\%. 0 denotes the HJ. The last column indicates the end time for
each simulation. Note that for the fiducial models (runs Aa1--3 and Ba1--5), results are shown for 
$t = 5 \times 10^8 T_{\rm K}$ and $t = 10^9 T_{\rm K}$.}
\label{tbl:results}
\end{deluxetable*}

\end{document}